\documentclass[twocolumn]{aastex63}
\usepackage{hyperref}
\usepackage{amsmath,amstext}
\usepackage[T1]{fontenc}
\usepackage{graphicx}
\usepackage[figure,figure*]{hypcap}
\usepackage{multirow}
\usepackage{comment}
\usepackage{nicefrac}

\def\sun{$_{\odot}$}

\def\f28{${f}_{2-8{\rm keV}}$}

\def\sun{$_{\odot}$}

\def\ergscm2{erg s$^{-1}$ cm$^{-2}$}
\def\yr-1{yr$^{-1}$}

\def\a{$\&$}
\def\x{$\times$}

\def\w{$\omega$}

\def\deg{$^{\rm o}$}
\def\asec{\ifmmode^{\prime\prime}\else$^{\prime\prime}$\fi}

\hypersetup{linkcolor=magenta}

\submitjournal{ApJ}
\received{March 20, 2020}
\accepted{May 6, 2020}

\shorttitle{Imaging Search for a Kilonova Counterpart to GW190814 
} 
\shortauthors{Vieira {\it et al.}}

\begin{document}

\title{A Deep CFHT Optical Search for a Counterpart to the Possible Neutron Star -- Black Hole Merger GW190814 }

\correspondingauthor{Nicholas~Vieira}
\email{nicholas.vieira@mail.mcgill.ca}

\author[0000-0001-7815-7604]{Nicholas~Vieira}
\affil{McGill Space Institute and Department of Physics, McGill University, 3600 rue University, Montreal, Quebec, H3A 2T8, Canada}

\author[0000-0001-8665-5523]{John~J.~Ruan}
\affil{McGill Space Institute and Department of Physics, McGill University, 3600 rue University, Montreal, Quebec, H3A 2T8, Canada}

\author[0000-0001-6803-2138]{Daryl Haggard}
\affil{McGill Space Institute and Department of Physics, McGill University, 3600 rue University, Montreal, Quebec, H3A 2T8, Canada}
\affil{CIFAR Azrieli Global Scholar, Gravity \& the Extreme Universe Program, Canadian Institute for Advanced Research, 661 University Avenue,
Suite 505, Toronto, Ontario, M5G 1M1, Canada}

\author[0000-0001-7081-0082]{Maria~R.~Drout}
\affil{Department of Astronomy and Astrophysics, University of Toronto, 50 St. George St., Toronto, Ontario, M5S 3H4, Canada}

\author[0000-0002-3310-1946]{Melania~C.~Nynka}
\affil{Kavli Institute For Astrophysics and Space Research, Massachusetts Institute of Technology, 77 Massachusetts Avenue, 37-241 Cambridge, MA 02139, USA}

\author{Hope~Boyce}
\affil{McGill Space Institute and Department of Physics, McGill University, 3600 rue University, Montreal, Quebec, H3A 2T8, Canada}

\author{Kristine~Spekkens}
\affil{Department of Physics, Royal Military College of Canada, P.O. Box 17000, Station Forces, Kingston, Ontario, K7K 7B4, Canada}

\author[0000-0001-6189-7665]{Samar~Safi-Harb}
\affil{Department of Physics and Astronomy, University of Manitoba, 311 Allen (Physics) Building, Winnipeg, Manitoba, R3T 2N2, Canada}

\author[0000-0002-7667-0081]{Raymond~G.~Carlberg}
\affil{Department of Astronomy and Astrophysics, University of Toronto, 50 St. George St., Toronto, Ontario, M5S 3H4, Canada}

\author[0000-0003-4619-339X]{Rodrigo Fern{\'a}ndez}
\affil{Department of Physics, University of Alberta, Edmonton, Alberta, T6G 2E1, Canada}

\author{Anthony~L.~Piro}
\affil{The Observatories of the Carnegie Institution for Science, 813 Santa Barbara St., Pasadena, CA 91101, USA}

\author[0000-0002-1338-490X]{Niloufar~Afsariardchi}
\affil{Department of Astronomy and Astrophysics, University of Toronto, 50 St. George St., Toronto, Ontario, M5S 3H4, Canada}

\author[0000-0003-4200-5064]{Dae-Sik~Moon}
\affil{Department of Astronomy and Astrophysics, University of Toronto, 50 St. George St., Toronto, Ontario, M5S 3H4, Canada}

\begin{abstract}
We present a wide-field optical imaging search for electromagnetic counterparts to the likely neutron star -- black hole (NS-BH) merger GW190814/S190814bv. This compact binary merger was detected through gravitational waves by the LIGO/Virgo interferometers, with masses suggestive of a NS-BH merger. We imaged the LIGO/Virgo localization region using the MegaCam instrument on the Canada-France-Hawaii Telescope. We describe our hybrid observing strategy of both tiling and galaxy-targeted observations, as well as our image differencing and transient detection pipeline. Our observing campaign produced some of the deepest multi-band images of the region between 1.7 and~8.7 days post-merger, reaching a $5\sigma$ depth of $g>22.8$ (AB mag) at 1.7~days and $i>23.1$ and $i>23.9$ at 3.7 and 8.7~days, respectively. These observations cover a mean total integrated probability of 67.0\% of the localization region. We find no compelling candidate transient counterparts to this merger in our images, which suggests that either the lighter object was tidally disrupted inside of the BH's innermost stable circular orbit, the transient lies outside of the observed sky footprint, or the lighter object is a low-mass BH. We use $5\sigma$ source detection upper limits from our images in the NS-BH interpretation of this merger to constrain the mass of the kilonova ejecta to be $M_{\mathrm{ej}} \lesssim 0.015M_{\odot}$ for a `blue' ($\kappa=0.5\mathrm{~cm^2 g^{-1}}$) kilonova, and $M_{\mathrm{ej}} \lesssim 0.04M_{\odot}$ for a `red' ($\kappa=5-10\mathrm{~cm^2 g^{-1}}$) kilonova. Our observations emphasize the key role of large-aperture telescopes and wide-field imagers such as CFHT MegaCam in enabling deep searches for electromagnetic counterparts to gravitational wave events.

\end{abstract}

\section{Introduction}\label{sec:intro}
On 14 August 2019 at 21:10:39.013 UTC, the Laser Interferometer Gravitational-wave Observatory (LIGO) and Virgo interferometers detected a high-confidence gravitational wave (GW) chirp from a compact object merger event, GW190814/S190814bv (\citealt{lvc19a}). Initial low-latency modeling of the gravitational waveform by the LIGO/Virgo Collaboration (LVC) classified this event as a MassGap merger, in which the mass of the lighter object is between 3 and 5~$M_\odot$, with $>99$\% probability. Less than half a day later, further modeling of the gravitational waveform re-classified this event as a potential merger between a neutron star (NS) and black hole (BH), with $>99$\% probability and an exceptionally low false-alarm rate (FAR) of approximately 1 in $10^{25}$ years, making GW190814 the first robust detection of a potential NS-BH merger (\citealt{lvc19b}). As a 3-detector event, this merger was also exceptionally well-localized, with a 50\% localization region of area 4.8 deg$^2$ and a 90\% localization region of area 23.1 deg$^2$. The luminosity distance measured from the amplitude of the GWs was $d_L = 267 \pm 52$ Mpc. Events classified as NS-BH mergers by LIGO/Virgo are mergers in which the heavier object is $>5M_\odot$, and the lighter object is $<3M_\odot$. Since the maximum NS mass is unclear (the most massive NSs currently known are $\sim2M_\odot$, \citealt{demorest10, cromartie19}) and dependent on the unknown equation of state of dense nuclear matter, whether or not the lighter object in GW190814 was actually a NS or a BH is unclear. Nonetheless, the MassGap/NS-BH classification, low FAR, and excellent localization of GW190814 make it a landmark event.

Follow-up searches for an electromagnetic (EM) counterpart to GW190814 can potentially reveal the nature of the event. The tidal disruption of a NS by a BH prior to a merger can dynamically eject neutron-rich material from the system, if this disruption occurs outside of the innermost stable circular orbit (ISCO) of the BH (\citealt{lattimer74, bethe98, rosswog05, shibata08, metzger08, etienne09, foucart14, kawaguchi15, kawaguchi16, kyutoku15, fernandez17, kyutoku18, fernandez19}). Following the merger of the NS and BH, an accretion disk is formed around a remnant BH and a tidal tail of both bound and unbound ejecta develops (\citealt{fernandez13, metzger14, just15, fernandez17, siegel17, foucart18, ruiz18, siegel18, christie19, fernandez19, foucart19}). The accretion disk and dynamical ejecta are dominated by radioactive isotopes synthesized via rapid capture of free neutrons, i.e., the $r$-process. These $r$-process isotopes radioactively decay, and their decay products undergo thermalization to power a transient kilonova observable at ultraviolet (UV), optical, and infrared (IR) wavelengths (\citealt{eichler89, li98, freiburghaus99, metzger10, kasen13, barnes13, tanaka13, tanaka14, just15, barnes16, kawaguchi16, tanaka18, tanaka19}). Tidal disruption outside of the ISCO is most likely in NS-BH systems with small binary mass ratios $q = \frac{M_{\mathrm{BH}}}{M_{\mathrm{NS}}}$ (e.g.  $M_{\mathrm{BH}}\lesssim8~M_\odot$, \citealt{shibata08, lovelace13, foucart14, foucart18, foucart19}) and/or highly-spinning black holes (e.g. $\chi_{\mathrm{BH}}\gtrsim0.7$, \citealt{etienne09, lovelace13, foucart14, kawaguchi15, foucart18}). Although the exact masses and spins of the merging compact objects in GW190814 have not yet been announced by the LVC, the detection of a kilonova counterpart in follow-up observations would confirm that the lighter object in this merger was indeed a NS. In addition, kilonovae remain an elusive class of transients in need of further study. The UV/optical/IR emission associated with the landmark NS-NS merger GW170817 represents the only unambiguous discovery of a kilonova to date (\citealt{abbottLIGO17a}). In that merger, the combination of excellent LIGO/Virgo GW localization (90\% region area of $31~\mathrm{deg^2}$, \citealt{abbottLIGO17a}), the Fermi-GBM/INTEGRAL detection of the short gamma-ray burst GRB170817A $\sim${}$1.7~\mathrm{s}$ after the GWs (\citealt{abbottLIGO17b}), and world-wide follow-up efforts led to the rapid localization of the EM counterpart to the galaxy NGC4993 at a distance of $\sim${}$40~\mathrm{Mpc}$, producing an unprecedented quantity of photometric and spectroscopic data. The GW170817 kilonova broadly matched theoretical predictions, revealing early emission  ($\lesssim1~\mathrm{day}$) that peaked in the UV/optical, followed by rapid reddening over the subsequent several days towards the IR (\citealt{andreoni17, arcavi17, coulter17, diaz17, drout17, evans17, hu17, kasliwal17, lipunov17, pian17, pozanenko17, shappee17, smartt17, tanvir17, troja17, utsumi17, valenti17}). However, due to a delay in the release of a three-detector localization map for the event, the first $\sim${}$10~\mathrm{hours}$ of the GW170817 kilonova were not observed by ground-based facilities. Furthermore, many questions on the late-time thermalization in the ejecta, the abundance patterns and importance of different radioactive isotopes in powering the kilonova, and the evolution of the opacity of the ejecta remain unanswered (e.g. \citealt{kasen19, khatami19, tanaka19}). Discoveries of new kilonovae are necessary to probe these questions, and the detection of a kilonova associated with a NS-BH merger in particular would be another landmark event. 

Here, we report results from a deep optical imaging search for a kilonova counterpart to GW190814 performed using the MegaCam instrument on the Canada-France-Hawaii Telescope (CFHT). Following the LIGO/Virgo detection of GW190814, we triggered our target-of-opportunity program (PI: Ruan) on the CFHT's MegaCam instrument to search for a kilonova-like EM counterpart in the GW localization region using wide-field optical imaging. Many other searches for possible counterparts to GW190814 were also conducted, including targeted imaging of individual galaxies in the localization region using telescopes with small fields of view (\citealt{gomez19, ackley20}) and tiled imaging of the full localization region using telescopes with large fields of view (optical/IR in \citealt{andreoni20, ackley20, watson20}; radio in \citealt{dobie19}). We took advantage of the wide $\sim1~\mathrm{deg^2}$ field of view of MegaCam, which enabled us to tile the full 50\% localization region, and to target individual galaxies in the larger 90\% localization region based on a prioritization scheme. To evaluate our observations, we take the median 5$\sigma$ source detection upper limit across all fields for a given epoch as the depth of our images. Our observations reach a depth of $g>22.8$ (AB mag) at 1.7 days post-merger and depths of $i>23.1$ and $i>23.9$ at 3.7 and 8.7 days post-merger, respectively. With these depths and our multi-band coverage of a large fraction of the localization region, these observations are among the most constraining.

The outline of this paper is as follows: In Section~\ref{sec:methods}, we describe our observations as well as our image differencing, transient detection, and transient vetting techniques. In Section~\ref{sec:other-searches}, we compare our kilonova search results to other searches reported to date. In Section~\ref{sec:models}, we use the results of our search and a simple model to constrain the presumed kilonova and the parameters of the merger ejecta. We summarize and conclude in Section~\ref{sec:conclusions}. Throughout this work, we assume a standard flat $\Lambda$CDM cosmology with $\Omega_\mathrm{m} = 0.309$, $\Omega_\Lambda = 0.691$, and $H_0 = 67.7$ km s$^{-1}$ Mpc$^{-1}$ (\citealt{planck16}).\newline

\begin{figure*} [t!]
\center{
\includegraphics[scale=0.59,angle=0]{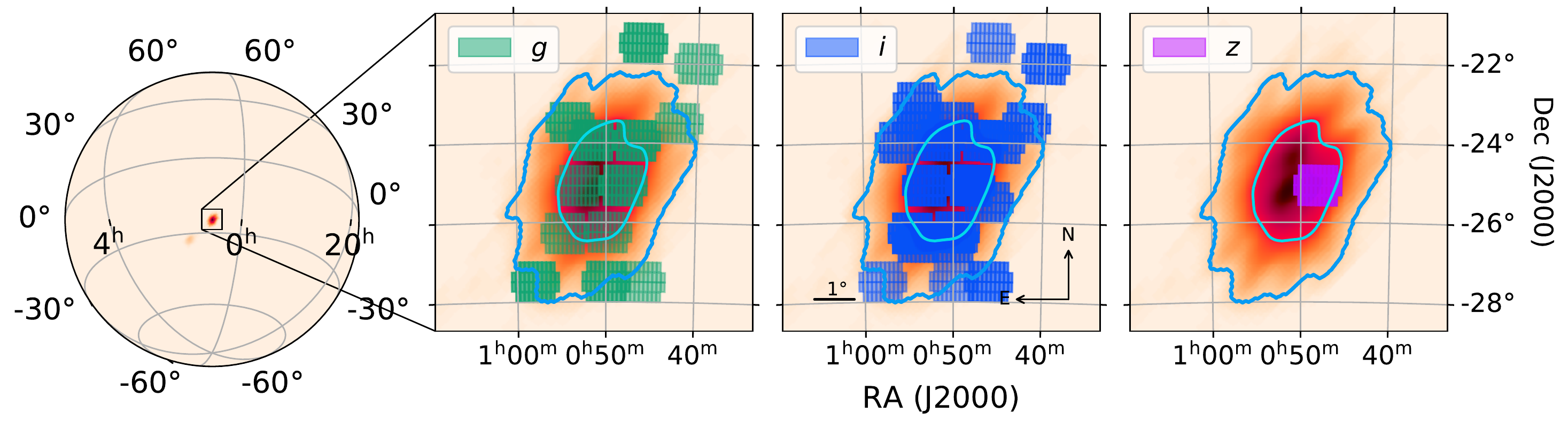}
}
\figcaption{\textbf{CFHT MegaCam imaging of the LIGO/Virgo GW190814 localization region in the $g$, $i$, and $z$ bands.} $g$-band imaging was obtained in the earliest observations following the merger, followed by the $i$ and $z$ bands. The inner cyan contours denote the 50\% localization region (area of 4.8~deg$^2$) and the outer darker blue contours denote the 90\% localization region (area of 23.1~deg$^2$). $z$-band images span a single field where we imaged a particular source of interest for 2 nights before ruling it out as a counterpart to the GW event. We neglect a secondary low-probability lobe of the region South-East of the primary lobe shown here. The gaps in our 50\% region pointings are due to an error during planning of observations.
}\label{fig:coverage}
\end{figure*}

\begin{deluxetable*}{ccccccc}
\centering
\tablecaption{{\bf CFHT MegaCam coverage of the LIGO/Virgo GW190814 localization region.} Columns include the band, time post-merger, areal coverage in the 50\% localization region, areal coverage in the 50\% $< p <$~90\% localization region, total integrated probability, the source of the reference images used in subsequent image differencing, and the median 5$\sigma$ limiting magnitude across all observed fields for the given observation epoch. Details of the pointings are included in Table~\ref{tab:details} (Appendix~\ref{app:obs}). Observations at 2.6 days post-merger were ended early due to poor observing conditions. $i$- and $z$-band images at 6.6 and 7.6 days post-merger were centered on a source of interest which was later ruled out as a counterpart to GW190814.
}
\tablehead{  
\colhead{band} & \colhead{$t-t_\mathrm{merger}$} & \colhead{50\% localization} & \colhead{50\% $< p <$~90\% localization} & \colhead{total integrated} & \colhead{reference} & \colhead{median 5$\sigma$} \vspace{-2.5pt} \\
\colhead{} & \colhead{[days]} & \colhead{region coverage} & \colhead{region coverage} & \colhead{probability} & \colhead{images} & \colhead{limiting magnitude}
 }
\startdata
 \vspace{2pt}
$g$ & 1.7 & 92.2\% & 31.5\% & 65.5\% & PS1 3$\pi$ & 22.8\\
 \vspace{2pt}
$g$ & 6.6 & 49.2\% & 19.5\% & 35.9\% & PS1 3$\pi$ & 23.6\\
 \vspace{2pt}
$i$ & 2.6 & 1.7\% & 10.3\% & 5.2\% & PS1 3$\pi$ & 21.8\\
 \vspace{2pt}
$i$ & 3.7 & 92.2\% & 31.9\% & 65.8\% (61.5\%)\tablenotemark{a} & MegaCam/PS1 3$\pi$ & 22.6 (23.1)\tablenotemark{a}\\
 \vspace{2pt}
$i$ & 4.7 & 94.0\% & 36.9\% & 70.5\% & MegaCam & 22.8\\
 \vspace{2pt}
$i$ & 6.6 & 22.7\% & 0.9\% & 13.0\% & MegaCam & 23.6\\
 \vspace{2pt}
$i$ & 7.6 & 22.7\% & 0.9\% & 13.0\% & MegaCam & 24.0\\
 \vspace{2pt}
$i$ & 8.7 & 94.0\% & 36.9\% & 70.5\% & MegaCam & 23.9\\
 \vspace{2pt}
$i$ & 20.5 & 94.0\% & 36.9\% & 70.5\% & \textit{N/A} & 23.8\\
 \vspace{2pt}
$z$ & 6.6 & 22.7\% & 0.9\% & 13.0\% & DECaLS & 22.7\\
 \vspace{2pt}
$z$ & 7.6 & 22.7\% & 0.9\% & 13.0\% & DECaLS & 23.1\\
\enddata
\tablenotetext{a}{Observing conditions at 3.7 days post-merger steadily improved over the course of the night. The numbers in parentheses are the integrated probability and median $5\sigma$ depth  which is attained if we exclude these earlier, poorer-quality observations. The total integrated probability is smaller, but the limiting magnitude is significantly deeper if these poorer-quality observations are cut.}
\end{deluxetable*}\label{tab:coverage}

\section{CFHT Follow-up Imaging of GW190814}\label{sec:methods}

\subsection{Details of the Observations}\label{ssc:obs}

The CFHT is a 3.58m aperture telescope, and MegaCam is a $\sim1~\mathrm{deg^{2}}$ field of view camera with a pixel scale of 0.185" per pixel. We used MegaCam to obtain $g$-, $i$-, and $z$-band imaging in the localization region of GW190814.  Due to the small LIGO/Virgo localization region of GW190814, we used a hybrid strategy of tiling the 50\% localization region, and galaxy-targeted observations in the 50\% $< p <$~90\% localization region. In the 50\% localization region (area of 4.8~deg$^2$), we used the large 1~deg$^2$ field of view of MegaCam to tile the region using 6 fields. However, tiling the larger 90\% localization region (area of 23.1~deg$^2$) was not feasible because the requisite observations were longer than the $\sim$4 hours per night during which this region of the sky was at sufficiently low airmass. Thus, we targeted individual galaxies in the 50\% $< p <$~90\% localization region to search for a counterpart using a galaxy prioritization scheme. 
In the 50\% $< p <90$\% localization region, galaxies were selected from the \textit{Galaxy List for the Advanced Detector Era} (GLADE, \citealt{dalya18})\footnote{\href{http://glade.elte.hu/}{glade.elte.hu/}} and prioritized based on their $B$-band luminosities (a tracer for the stellar mass of the galaxy), photometric redshifts, and positions in the localization region. 

To determine the photometric depths and exposure times required for our search, we scaled the peak apparent magnitude of the GW170817 kilonova ($i\sim17.5$ at $\sim1$ day post-merger; see Section~\ref{sec:intro} for full list of citations) to the mean distance $d_L = 267$ Mpc of GW190814, yielding a scaled peak of $i\sim21.5$ (absolute magnitude $M_i\sim-15.6$). Numerical relativity simulations of NS-BH mergers find similar peak magnitudes such as $M_i\sim-15.0$ ($i\sim22.1$) at $\sim3$ days (\citealt{kawaguchi16}) and $M_i\sim-15.6$ ($i\sim21.5$) at $\sim2$ days (\citealt{tanaka18}). A depth of $i\sim22$ was therefore taken to be sufficient, and exposure times were computed to enable MegaCam to reach this depth. We obtained 5 $\times$ 300 s exposures and 5 $\times$ 200 s exposures for images in the 50\% and 50\% $< p <$~90\% regions, respectively. We used standard dithering patterns to enable cosmic-ray rejection and to fill chip gaps in the MegaCam CCD array. The positions of all exposures in each filter are shown in Figure~\ref{fig:coverage}, overlaid on the \texttt{LALInference} LIGO/Virgo localization region (\citealt{lvc19b})\footnote{\href{https://gracedb.ligo.org/superevents/S190814bv/view/}{gracedb.ligo.org/superevents/S190814bv/view/}}. The median seeing across all images was $\sim0.80$". Our observations began at 1.7 days post-merger in the $g$-band and continued from 3.7 days to 8.7 days in the $i$-band. This strategy was based on observations of the GW170817 kilonova, which peaked first in the UV on timescales of $\lesssim1~\mathrm{day}$ before becoming redder and peaking in the optical/IR over the next $\sim$ 10 days. We also acquired $z$-band images at 6.6 and 7.6 days while following a particular source of interest which was later disqualified as a candidate counterpart to GW190814. Finally, we also acquired $i$-band images at 20.5 days to supplement archival reference images from other surveys for use in our image differencing (Section~\ref{ssc:imagedifferencing}). Details of each of these observations, including the areal coverage fraction for each epoch and the total integrated probability, are listed in Table~\ref{tab:coverage}. A full observation log is given Table~\ref{tab:details} (Appendix~\ref{app:obs}). Areal coverage fractions and integrated probabilities were computed using Multi-Order Coverage maps via \href{https://github.com/cds-astro/mocpy/}{\texttt{MOCpy}} (\citealt{fernique14}) and \href{https://lscsoft.docs.ligo.org/ligo.skymap/}{\texttt{ligo.skymap}}. 

\begin{figure*}[ht!]
\center{
\includegraphics[scale=0.52,angle=0]{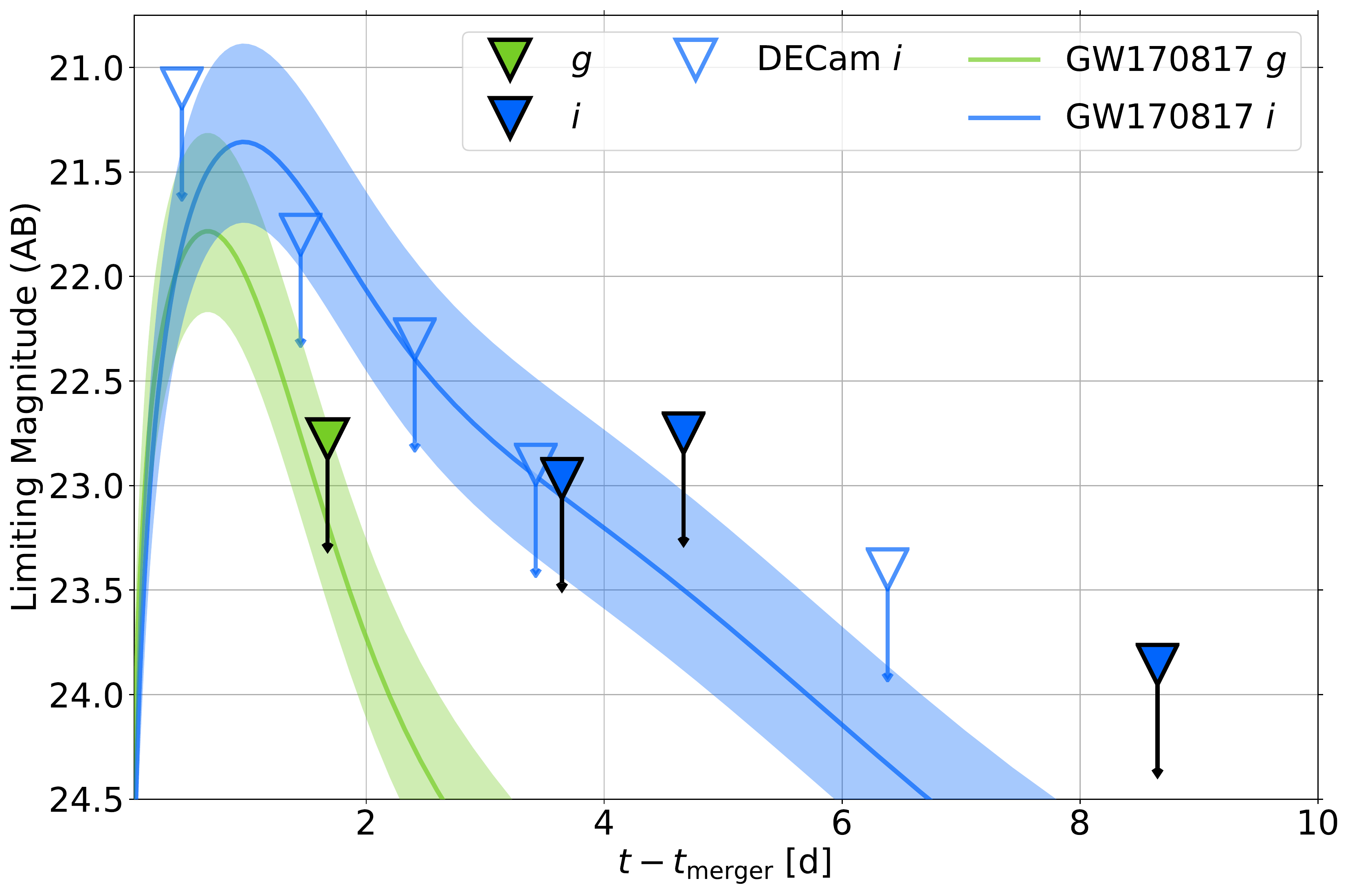}
}
\figcaption{\textbf{5$\sigma$ CFHT MegaCam limiting magnitudes achieved for selected epochs of imaging.} The depths of the earliest images were affected by the brightness of the moon, which waned in the following days. All of the limits shown stem from observations acquired with $>90\%$ areal coverage of the 50\% region (see Table~\ref{tab:coverage}). For comparison, we show the depths achieved by DECam in the $i$-band (\citealt{andreoni20}). We also show model light curves of the EM counterpart to GW170817, constructed using the best-fit 3-component symmetric kilonova model of \texttt{MOSFiT} presented in \citet{villar17} and re-scaled to the distance $d_L = 267 \pm 52$ Mpc of GW190814. The upper limits imposed by MegaCam are among the strictest achieved limits for GW190814. A GW170817-like kilonova at a distance of 215-267 Mpc would likely have been detected in our MegaCam observations.}
\label{fig:depth}
\end{figure*}

The $5\sigma$ depths for each epoch of observation are presented in Figure~\ref{fig:depth}. These depths were computed by performing aperture photometry on a background-only region of each MegaCam CCD image to determine the minimum flux required for a source to be detected at $\geqslant5\sigma$ significance. The depths shown for each epoch are the median depths achieved over the course of the observations and across the observed field. These depths vary on the order of $\sim0.1-0.2$ mag across the observed field for a given epoch. We also show the depths achieved by DECam (\citealt{andreoni20}), for comparison. Finally, we show model light curves for the EM counterpart to GW170817, constructed using the best-fit parameters of the 3-component symmetric kilonova model of \texttt{MOSFiT}\footnote{\href{https://github.com/guillochon/MOSFiT}{github.com/guillochon/MOSFiT}} (\citealt{guillochon18}) presented in \cite{villar17} and re-scaled to the distance $d_L = 267 \pm 52$ Mpc of GW190814. The $K$-corrections to the $g$- and $i$-band light curves of GW170817 at this distance, for the colours given in \cite{villar17}, are of the order $|K|<0.05$ near peak magnitude and are neglected. The upper limits obtained by MegaCam are among the most constraining for this merger, and indicate that we would have likely detected a GW170817-like kilonova at a distance of 215-267 Mpc. 

\subsection{Data Reduction}
\label{ssc:datareduction}

The MegaCam images were reduced using the CFHT's \texttt{Elixir}\footnote{\href{https://www.cfht.hawaii.edu/Instruments/Elixir/}{cfht.hawaii.edu/Instruments/Elixir/}} pipeline, which includes bias, dark, flat-field, and fringe-frame corrections.
For each exposure, we perform astrometric calibration using \texttt{astrometry.net} \citep{lang10} based on the Gaia DR2 catalog \citep{gaia18, lindegren18}. We then coadd the 5 exposures for each field on each night and perform photometric calibration based on the Pan-STARRS1 3$\pi$ survey \citep[PS1,][]{chambers16}.

\subsection{Image Differencing and Transient Search}
\label{ssc:imagedifferencing}

To search for possible kilonovae in our multi-band MegaCam images, we perform image differencing. For reference images in the $g$- and $z$- bands, we use archival images from PS1 and the Dark Energy Camera (DECam) Legacy Survey \citep[DECaLS,][]{abbott18, dey19}, respectively. PS1 is selected for the $g$-band because it is sufficiently deep to be used as a reference given the depth we achieve with MegaCam. Likewise, the depths achieved by DECaLS in the $z$-band are also sufficient for this purpose. PS1 and DECaLS images were obtained approximately 3~and 2~years before GW190814, respectively. 

Because our $i$-band images are deeper than those of PS1 and the DECaLS archive does not include the $i$-band, we use additional deep MegaCam observations obtained at $\sim$20.5~days as reference images. For GW170817, the $i$-band emission faded by $>$4 mags at 7 days post-merger (\citealt{siebert17}), and similar fading is expected in the case of a NS-BH kilonova (e.g. \citealt{kawaguchi16, fernandez17, tanaka18}). We therefore expect any kilonova emission in the $i$-band to have faded significantly at 20.5 days, thus enabling detection of earlier kilonova transients. For a small number of $i$-band images which all lie in the 50\% $< p <$~90\% localization region ($<$20\% of all $i$-band data), we did not acquire MegaCam reference images, and use PS1 instead. 

To enable image differencing between the science and reference images, we perform image alignment, background subtraction, and saturated/bad pixel masking. We extract sources from the images using image segmentation via the \texttt{photutils}\footnote{\href{https://photutils.readthedocs.io/en/stable/}{photutils.readthedocs.io}} package (\citealt{photutils19}). We then match point sources in the science and reference images to compute the affine transformation which aligns the science image with the reference image using the \texttt{astroalign}\footnote{\href{https://github.com/toros-astro/astroalign}{github.com/toros-astro/astroalign}} package (\citealt{beroiz19}). Sources are re-extracted in the aligned images and masked to compute a smoothly-varying median background which is then subtracted from each image. Finally, we flag saturated sources and/or bad pixels which should be masked in subsequent image differencing. We perform the described background subtraction and saturated/bad pixel masking for both science and reference images.

We use the High Order Transform of Point-Spread Function (PSF) And Template Subtraction\footnote{\href{https://github.com/acbecker/hotpants}{github.com/acbecker/hotpants}} software (\citealt{becker15}) to perform our image differencing. We adopt a set of 3 Gaussian $\times$ polynomial functions as the basis for our convolution kernel. The Gaussian terms have FWHM approximately equal to half of the median seeing ($\sim0.4$"), the seeing ($\sim0.8$"), and twice the size of the seeing ($\sim1.6$"), respectively. 
Each science and reference image is divided into 10 $\times$ 10 subregions, and a convolution kernel is fit to centroids present in both the science and reference image in each of these 100 subregions. This produces a convolution kernel which varies across the image, accounting for small variations in the PSF, which is then used to match the PSF of the science and reference images. Finally, the reference image is subtracted from the science image, and the final difference image is normalized to the photometric system of the science image. We assume no additional spatial variations in the background (which has already been subtracted) or convolution kernel across the image. 

We again use \texttt{photutils} to search for sources in our difference images. The detection significance for each source is set using the standard deviation of the good pixels in the difference image, which is uniform across each image. We only accept sources detected at $\geqslant${}$5\sigma$ significance. For each such source, we obtain its outermost isophote (i.e, the largest isophote within which all pixels are connected and are at $\geqslant${}$5\sigma$ above the background). Sources are rejected as being likely spurious if they fall into any of the following categories:
\begin{enumerate}
  \item Sources with isophotal pixel area $<$20~$\mathrm{pix}^2$. A circular source with $\mathrm{area}~ 20~\mathrm{{pix}^2}$ would have a radius of $0.5$", below the typical seeing ($\sim0.7-0.9$") in the images.
  \item Sources that display a `dipole' pattern in the difference image. We cross-match all positive and negative $>${}$5\sigma$ sources in each difference image, and select pairs of positive and negative sources lying within 2.0" of each other. We then reject matched pairs for which the flux ratio of the brighter source over the dimmer source is $<$5 (i.e., both parts of the dipole are similar in brightness). Such sources are a result of misalignment of the images or difficulty in matching the PSF of the science and reference images, especially near extended sources. 
  \item Sources with isophotal axial ratio, i.e. elongation, $>$ 2.0. These elongated sources are most likely image artifacts, residuals from bad subtractions in the vicinity of bright or extended objects, or cosmic rays.
  \item Sources with pixel area $>${}$400~\mathrm{pix}^2$. A circular source with $\mathrm{area}~400~\mathrm{{pix}^2}$ would have a radius of $2.1$". These large sources are most likely subtraction artifacts in the vicinity of bright, saturated sources which were not completely masked prior to image differencing.
\end{enumerate}
Finally, if the number of sources in a single MegaCam CCD image that pass these criteria exceeds 50, the difference image is likely to be of poor quality and thus we reject the entire image altogether. This occurs for $<0.1$\% of all images. For MegaCam-subtracted $i$-band images, 15,828 sources pass all of the above rejection criteria. For PS1-subtracted $g$-band images and DECaLS-subtracted $z$-band images, we obtain 2,379 and 1,190 sources, respectively. For the small number of PS1-subtracted $i$-band images, we obtain 1,986 sources. This yields a total of 15,828 + 2,379 + 1,190 + 1,986 = 21,383 candidate transient counterparts which must be classified. We produce $63 \times 63$ pixel cutouts of the science, reference, and difference images centered on each source (henceforth referred to as `triplets') for further vetting.

\begin{figure*} [t!]
\center{
\includegraphics[scale=0.17,angle=0]{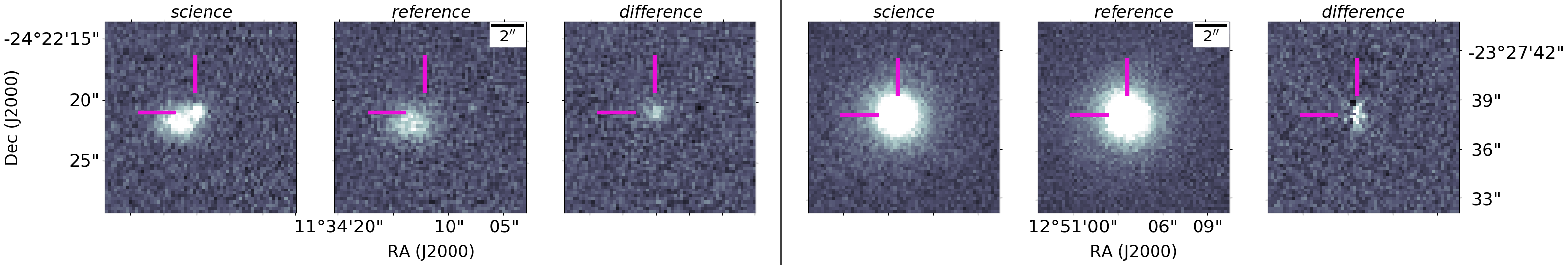}
}
\figcaption{\textbf{Example `triplets' of a real (left) and bogus (right) source.} Each triplet shows the science, reference, and difference images cutout and centered on the candidate of interest. The left triplet here is centered on AT2019nxe, a known transient, and is therefore real. The right triplet's difference image shows residuals from subtraction in the vicinity of a bright, large object and would therefore be classified as bogus by visual inspection. Pink crosshairs denote the location of the difference image peak. The vast majority of sources which pass the initial rejection criteria are bogus. A real-bogus algorithm is used to minimize the number of sources which require visual inspection.}\label{fig:triplet_examples}
\end{figure*}

\subsection{Vetting of Candidate Counterparts}
\label{ssc:braai}

The vast majority of the 21,383 candidate transient counterparts that pass the initial rejection criteria are spurious. This issue is common to all transient-detection pipelines, and additional vetting using `real-bogus' algorithms (a form of binary classification) based on machine-learning techniques is a common solution. We use the Bogus-Real Adversarial Artificial Intelligence (\texttt{braai})\footnote{\href{https://github.com/dmitryduev/braai}{github.com/dmitryduev/braai}} software package, which enables training and use of the \textit{VGG6} convolutional neural network via the high-level \texttt{TensorFlow} software (\citealt{duev19}). This neural network accepts as input the triplets of candidate counterparts and outputs a Real-Bogus (RB) score from 0 (definite bogus) to 1 (definite real) for each. 

To train a neural network for application to CFHT MegaCam data, we built a training set of 1,582 randomly-selected sources from our MegaCam-subtracted $i$-band images and 402 randomly-selected sources from our PS1-subtracted $g$-band images. These represent $\approx$10\% of our dataset altogether. Five team members then independently visually inspected each triplet and assigned them a label of 0 or 1. We then averaged the results across all five inspectors and rounded down to 0 or up to 1. Example triplets for a real and bogus source are shown in Figure~\ref{fig:triplet_examples}. 

From our visual inspection, only 116/1582 $=7.3\%$ sources in the $i$-band and 25/402 $=6.2\%$ sources in the $g$-band were identified as potentially `real' by eye. A low fraction of real sources at this stage is expected, but presents a challenge for training the neural network to classify new sources. We thus supplemented our training set by selecting additional sources from our 21,383 candidate transient counterparts which correspond to known transient sources. Specifically, we cross-matched all 21,383 candidate counterparts, spanning all epochs and bands, with sources from the Transient Name Server (TNS)\footnote{\href{https://wis-tns.weizmann.ac.il/}{wis-tns.weizmann.ac.il/} \label{fn:TNS}}. By selecting matches with good subtractions in our images, we added an additional 68 triplets to our training set (39 in the $i$-band with MegaCam templates, 29 in the $g$- and $i$-bands with PS1 templates), representing 26 distinct sources. Finally, we augmented the number of real sources in our training set by rotating the cutout science, reference, and difference image of only the real triplets in increments of 90\deg, effectively quadrupling the number of real sources in the dataset. 

As a result, the final dataset used to train the neural network has an approximate $836/1843 = 31\%/69\%$ real/bogus ratio. To mitigate the remaining class imbalance, a reduced weight equal to the ratio of real/bogus triplets is applied to the loss computed for the bogus triplets during training. The effect of this weighting is to make the loss of the model (which is to be minimized as the model converges to greater accuracy) more sensitive to misclassification of real sources than bogus sources. We use a training/validation/test set split of $81\%/9\%/10\%$ to train our model.

\begin{figure} [!t]
\center{
\includegraphics[scale=0.35,angle=0]{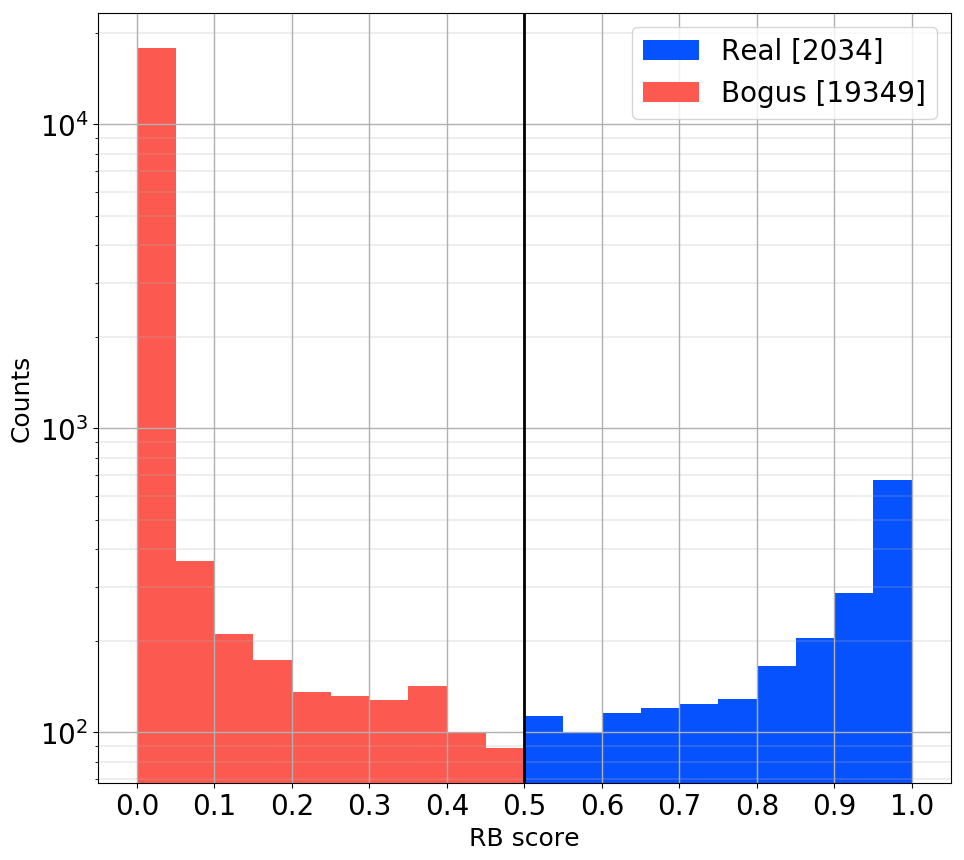}
}
\figcaption{\textbf{Histogram of the Real-Bogus (RB) scores assigned to each candidate by the trained model.} Of 21,383 candidates, 2,034 (9.5\%) are classified as real for RB $\geqslant$ 0.5.}
\label{fig:RB_histo}
\end{figure}

Although we use a relatively small and imbalanced dataset for training, we obtain an accurate and useful model that is sufficiently sensitive to detect the real sources in our test set. For a score RB $\geqslant0.5$ denoting a `real' source, the model we use yields a false positive rate (FPR) of 4.6\% and false negative rate (FNR) of 4.3\% when applied to the test set, yielding a mean misclassification error of 4.5\%. The model does not yet discard all spurious signals, but does effectively reduce the total number of candidate transients which require further analysis by an order of magnitude. The ability of the model to classify will improve with future follow-up campaigns and the addition of MegaCam data from other sources. Additional details on the performance of this MegaCam-tailored \texttt{braai} neural network are presented in Appendix~\ref{app:rb}.

A histogram of the RB scores assigned to all 21,383 candidates is shown in Figure~\ref{fig:RB_histo}. We are left with 1,462 candidates with RB $\geqslant$ 0.5 in MegaCam-subtracted $i$-band images, 259 in PS1-subtracted $g$-band images, 249 in PS1-subtracted $i$-band images, and 64 in DECaLS-subtracted $z$-band images, for a total of 2,034 candidates of interest. This translates to a 90.5\% decrease in the number of candidates which require further analysis. 

\begin{figure*} [!ht]
\center{
\includegraphics[scale=0.48,angle=0]{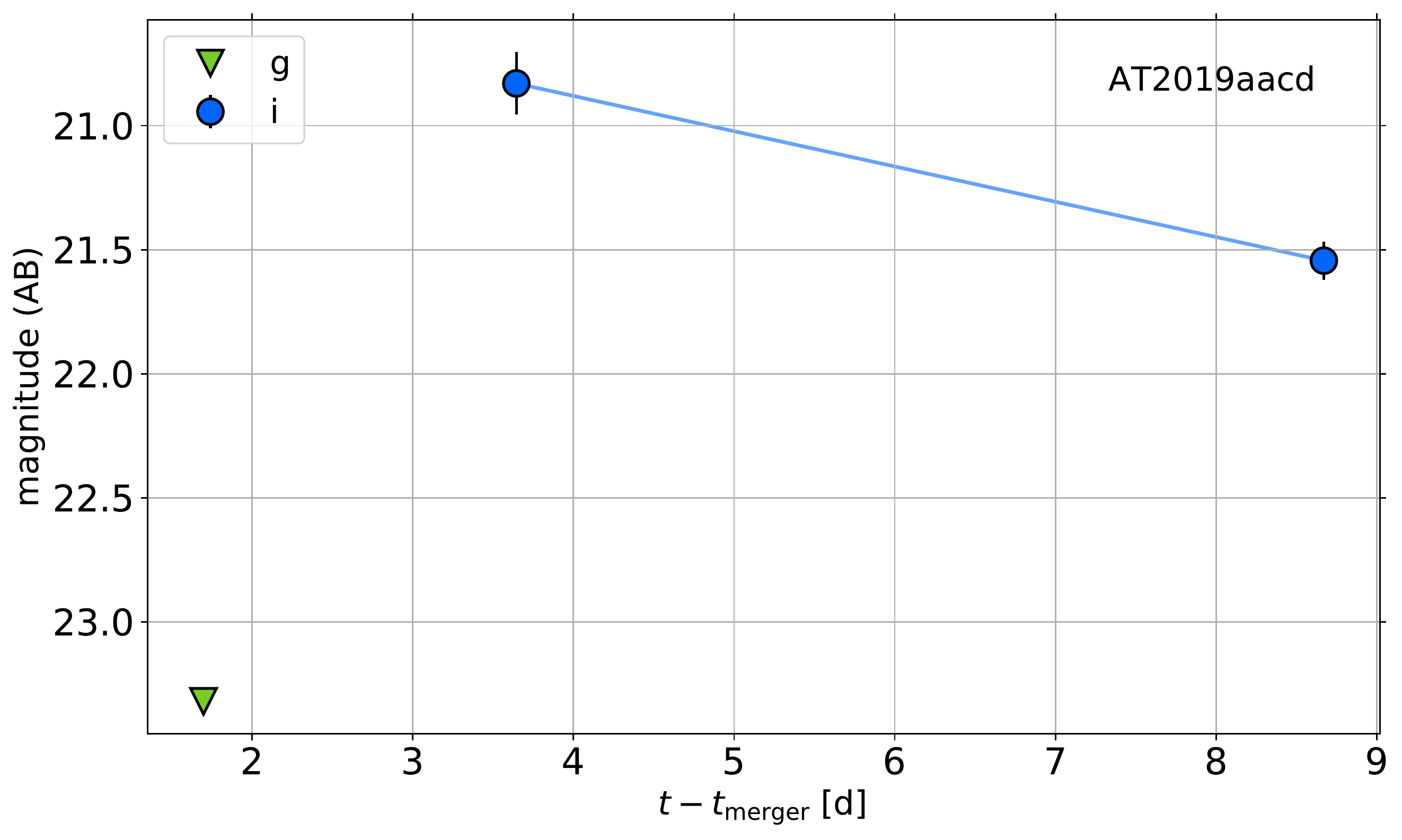}{\textbf{(a)}}
\includegraphics[scale=0.158,angle=0]{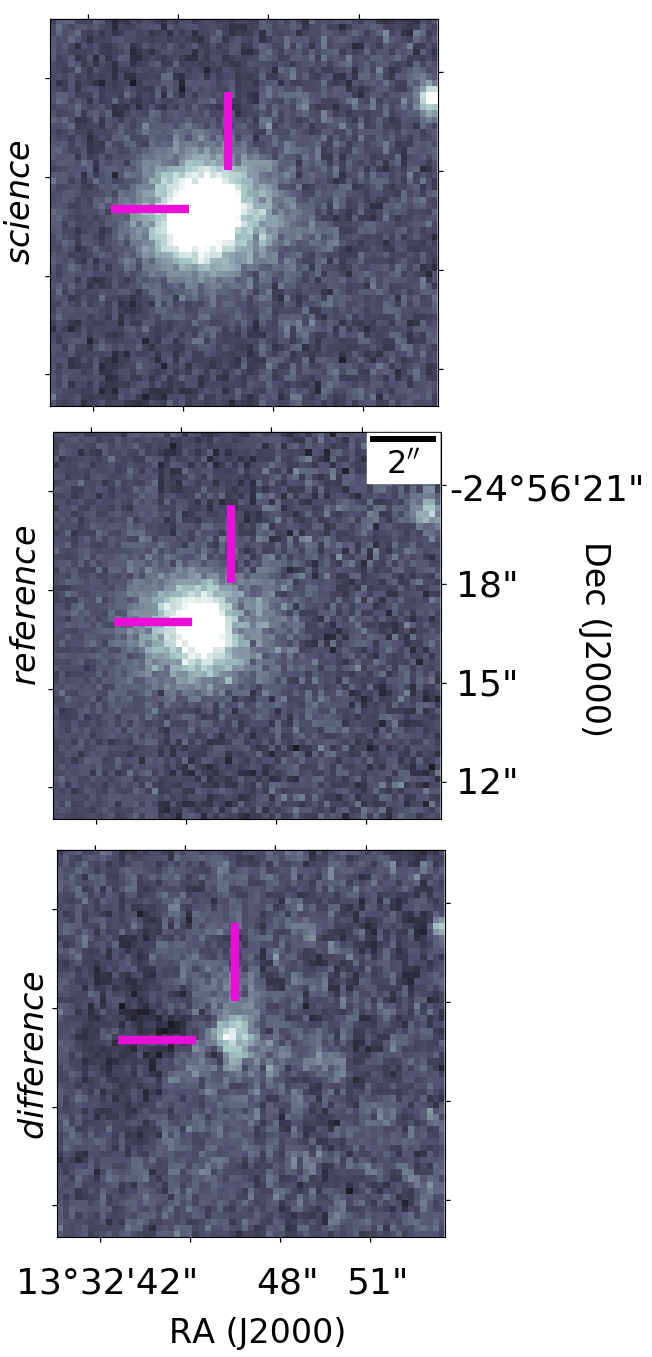}{\textbf{(b)}}
}
\figcaption{\textbf{Light curve (a) and corresponding triplet (b) for the new CFHT MegaCam transient AT2019aacd.} \textbf{(a)} Reliable photometry for AT2019aacd at 4.7 days was not obtainable due to host contamination and poorer seeing, and the source did not lie in the observation footprint at 6.6 days. \textbf{(b)} The reference image shown here is from MegaCam, and the triplet was acquired at 8.7 days. The pink crosshair denotes the location of the difference image peak. Examining the triplets in conjunction with the light curves provides a complete picture of the behaviour of a source. In particular, for distant sources such as GW190814 that may reside in an uncatalogued host galaxy, visually examining the triplets is necessary to determine whether or not a host is present. In this case, the source is offset by 1.38" (1.79 kpc at the distance $d_L = 267$ Mpc of GW190814) from the potential host galaxy WISEA J005411.20-245617.9, which is clearly visible in both the science and reference images.}
\label{fig:zau_LC}
\end{figure*}

\subsection{Inspection of Remaining Candidates}
\label{ssc:remain}
To further restrict the sample of the remaining 2,034 candidates to potentially real counterparts, each one is cross-matched with:
\begin{enumerate}
    \item Known transients from the Transient Name Server (TNS).
    \item Stellar (non-extended) objects in the PS1 3$\pi$ survey (\citealt{chambers16}).
    \item Known variable stars in the American Association of Variable Star Observers international Variable Star indeX (AAVSO-VSX, \citealt{watson06}).
    \item Quasars/active galactic nuclei from the V{\'e}ron-Cetty \& V{\'e}ron quasar/active nuclei catalogue (13th ed., \citealt{veron10}) and the Million Quasars (MILLIQUAS) catalogue (v6.3, updated 16 June 2019, \citealt{flesch19}).
\end{enumerate} 

We find 1 variable star from the AAVSO, and 61 quasars (representing 46 distinct objects). In addition, 67/68 triplets corresponding to known TNS sources are correctly recovered as real. The final sample of 1,905 candidates then only contains new transient sources and/or image artifacts. We visually inspect the remaining sources to remove false positives, and cross-match those which pass visual inspection against themselves to determine how many sources are multi-epoch/multi-band detections of the same intrinsic object. We are left with a total of 117 real triplets which represent 115 distinct sources. The fact that only $1+61+67+117=246$ of the 2,034 candidates either corresponded to known sources or passed the subsequent visual inspection implies a much higher FPR than the expected 4.6\%. This indicates that the classifier is over-fit to our training set, and that the 4.6\% FPR applies only to the training set and not to new data. We note, however, that the goal of the model at this stage is only to minimize the number of candidates which require visual inspection, as has been achieved with the 90.5\% reduction in the number of candidates in need of inspection. Nonetheless, the model will improve with a larger input training set, which will be available in the future. 

We perform aperture photometry at the coordinates of the 115 candidates of interest in the difference images for all available epochs and bands to produce light curves for each source. Examining the light curves for these sources in conjunction with the triplets at each epoch, we find a single transient source of interest that had not previously been reported to the TNS. The source is located at $\mathrm{RA}=13.54700, \mathrm{Dec}=-24.93824$. For all other non-reported sources, light curves and triplets suggests that they are either variable stars or image differencing artifacts. We assign this new transient the name CFHT0054-2456zau, subsequently reported to the Transient Name Server with identifier AT2019aacd. The light curve and a corresponding triplet of this source are shown in Figure~\ref{fig:zau_LC}.

In our observations, AT2019aacd is detected at $i = 20.83 \pm 0.13$ at 3.7 days post-merger. It then fades in the $i$-band by $0.71 \pm 0.15$ mag over 5 days ($\Delta i = 0.14 \pm 0.03~\mathrm{mag/day}$) to a magnitude of $i = 21.54\pm0.08$. Galactic extinction in the direction of this transient $E(B-V) \approx 0.02$, and is therefore neglected (\citealt{schlafly11}). The source is not detected to a $5\sigma$ limit of $g > 23.3$ at 1.7 days post-merger. This indicates the possibility of a very red transient, although the lack of simultaneous observations in multiple bands preclude definitive statements about transient color, rise time, or explosion epoch.

AT2019aacd is offset by 1.38" (1.79 kpc at the median GW190814 distance of $d_L = 267$ Mpc) from a potential host galaxy WISEA J005411.20-245617.9, which is clearly visible and extended in the CFHT science and reference images. This galaxy does not have a known redshift. The putative host has an $i$-band magnitude of $i = 18.8 \pm 0.1$ (corresponding to $M_i = -19.1$ or $\sim0.5 L_*$ at $d_L = 267$ Mpc). Following the methodology of \cite{bloom02} and \cite{berger10}, we find a probability of chance alignment between AT2019aacd and WISEA J005411.20-245617.9 of only  0.2\%, indicating very likely association.

The presence of a host, relatively rapid fading, and potential red colour of this new transient are intriguing. However, the observed brightness of $i = 20.83 \pm 0.13$ at 3.7 days corresponds to an absolute magnitude $M_i = -16.30 \pm 0.13$ at the median GW190814 distance of $d_L = 267$ Mpc. This is $\sim2.6$ mag brighter than GW170817 was at a similar epoch, and significantly brighter than expected for most kilonova models. In addition, the observed decline rate is marginally consistent with other classes of transients (e.g. Type Ic supernovae (SNe), \citealt{siebert17}). Further observations might have revealed the nature of the source. However, the transient detection pipeline we have described here was unfortunately not yet complete at the time of GW190814, and AT2019aacd was only discovered several weeks post-merger after this pipeline was completed. 

We thus conclude that we do not detect any high-confidence EM counterpart to GW190814 in our images. Since we detect no kilonova-like source, we therefore quantify and evaluate our coverage of the GW localization region in comparison with other searches, and employ a simple kilonova model to constrain the parameters of any possible merger ejecta using the upper limits derived from our deep CFHT observations.

\section{Comparison to Other Searches}\label{sec:other-searches}
\begin{figure*} [t!]
\center{
\includegraphics[scale=0.54,angle=0]{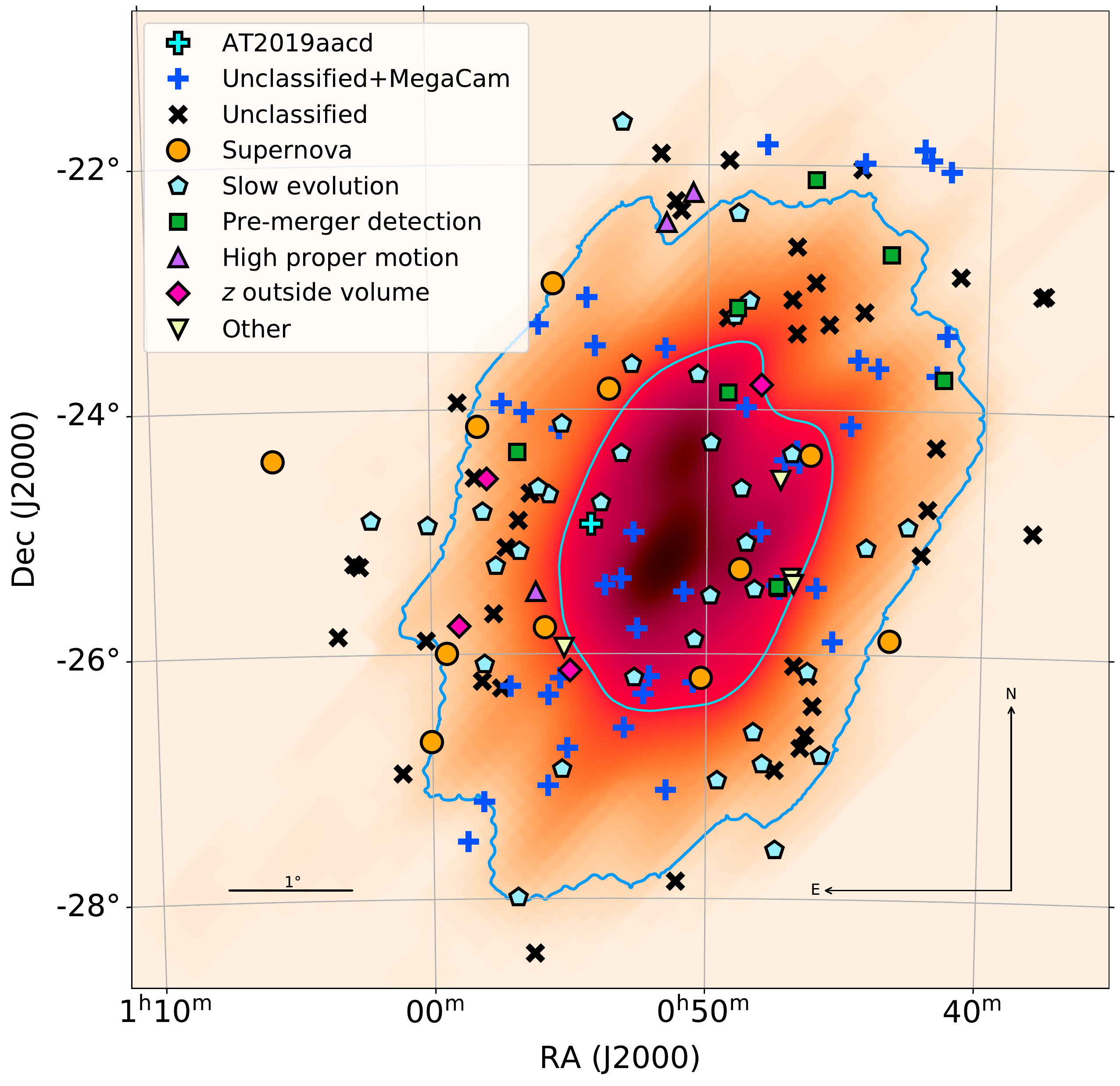}
}
\figcaption{
\textbf{All transients near the main lobe of the localization region which were reported to the Transient Name Server (TNS) in the 3 weeks immediately following the merger.} The sources deemed most promising were followed and classified. The most common reasons for transients being disqualified as candidate counterparts to the GW were spectroscopic classification as a supernova or photometric classification as a source evolving too slowly (e.g. $\Delta m < 0.1~\mathrm{mag/day}$) to correspond to the predicted kilonova. Other common reasons for disqualification were detection in archival pre-merger images, displaying a high proper motion consistent with being a Solar System object, or association with a host galaxy at redshift $z$ outside the LVC 2$\sigma$ confidence region. Sources classified as `Other' either displayed a featureless spectrum, were associated with the bright foreground galaxy NGC253, were consistent with being a nuclear source in some galaxy, or displayed no host at all. Sources labelled `Unclassified+MegaCam' lie in the MegaCam observation footprint, while those labelled `Unclassified' do not. All classifications were taken from \cite{ackley20}, \cite{andreoni20}, or directly from Gamma-ray Coordination Network Circulars (GCNs). See \cite{ackley20} in particular for a detailed summary of the classifications of TNS sources. We also show the location of the new CFHT source AT2019aacd highlighted in Section~\ref{ssc:remain}.}
\label{fig:reportedtransients}
\end{figure*}
\begin{figure*} [t!]
\center{
\includegraphics[scale=0.47,angle=0]{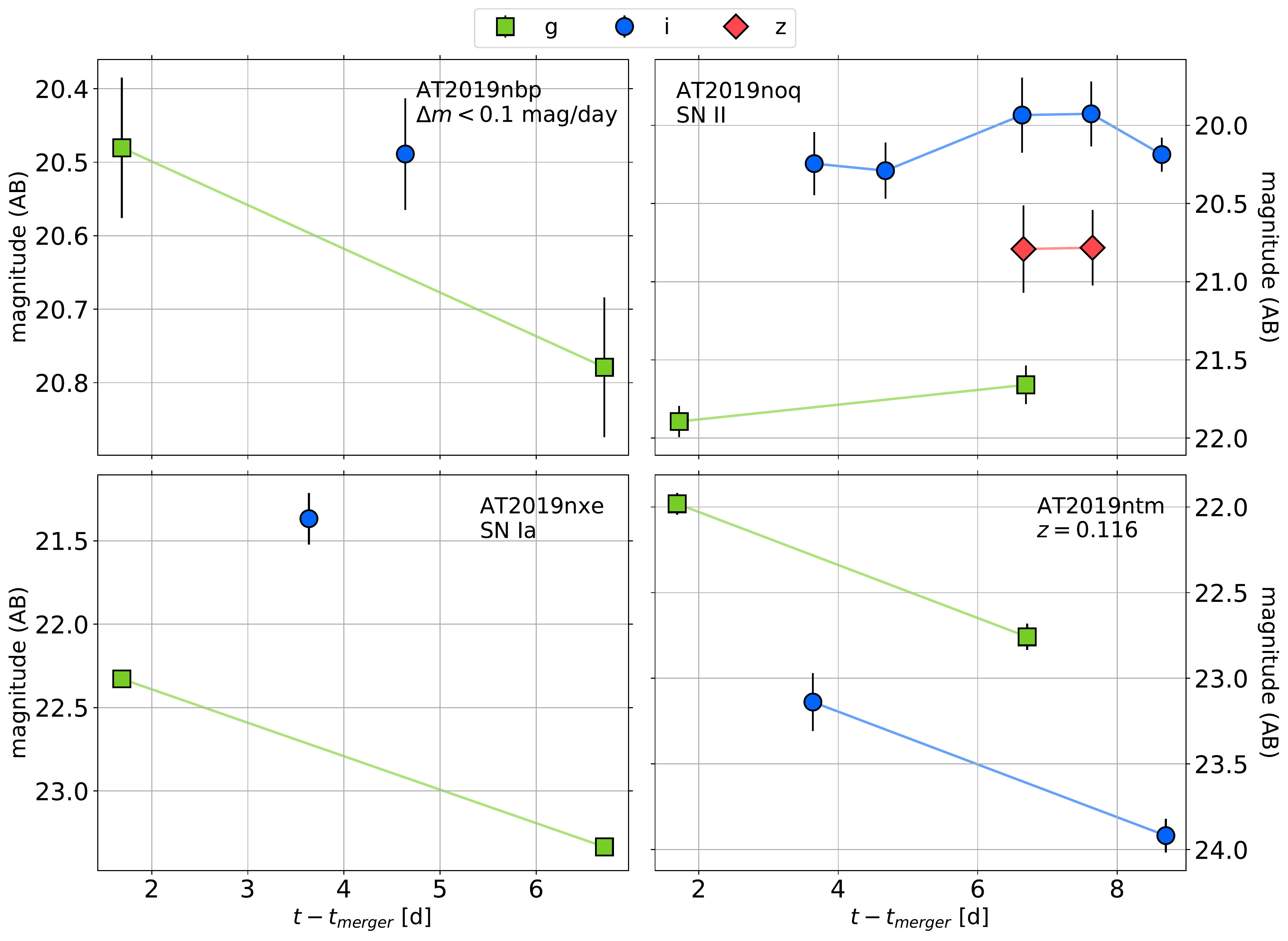}
}
\figcaption{\textbf{Light curves for 4 example sources which were reported to the TNS but eventually disqualified as counterparts.} AT2019nbp was associated with a potential host galaxy within the 2$\sigma$ localization volume of GW190814, but was later disqualified because its photometric evolution was inconsistent with the expected kilonova ($\Delta m < 0.1~\mathrm{mag/day}$, \citealt{andreoni20}). It was also detected in pre-merger images (\citealt{ackley20}). AT2019noq and AT2019nxe were classified in follow-up spectroscopy as Type II and Ia SNe, respectively (\citealt{andreoni20}). The host of AT2019ntm displayed a potential H$\alpha$ line (obtained by the William Herschel Telescope on 2019-09-09) which corresponded to $z=0.116$, outside the LVC $2\sigma$ confidence region of GW190814 (\citealt{ackley20}). Error bars on the $g$-band photometry for AT2019nxe are on the order of the point size.
}
\label{fig:sample_lightcurves}
\end{figure*}
Many other teams performed follow-up observations of GW190814 and reported dozens of possible counterparts, using a wide variety of telescopes and at many wavelengths. Candidates were reported via Gamma-ray Coordination Network Circulars (GCNs)\footnote{All GCNs associated with GW190814 are archived at \href{https://gcn.gsfc.nasa.gov/other/S190814bv.gcn3}{gcn.gsfc.nasa.gov/other/S190814bv.gcn3}.}. Figure~\ref{fig:reportedtransients} shows all transients in the main lobe of the localization region which were reported to the TNS from the time of the merger to 3 weeks post-merger. For many sources, the observed photometric evolution was too  slow to correspond to the predicted kilonova. Other sources were disqualified due to spectral classification which showed that they were most likely SNe. Of the reported transients in our footprint, we recover 26. There are multiple reasons why we do not recover all sources: (1) some sources evolve too slowly and do not present a detectable difference in flux over the $\leqslant20.5$ days between our $i$-band images and templates, which constitute most of our data, and (2) many sources were eventually found in archival pre-merger images, suggesting that they may be transients which evolve on timescales $\gg20.5$ days. In Figure~\ref{fig:sample_lightcurves}, we show example MegaCam light curves for four of these sources which appeared promising in the days following the merger but were later disqualified: AT2019nbp (slow photometric evolution $\Delta m < 0.1~\mathrm{mag/day}$,~\citealt{andreoni20}, and pre-merger detection,~\citealt{ackley20}), AT2019noq (SN Type-II,~\citealt{andreoni20}), AT2019nxe (SN Type-Ia,~\citealt{andreoni20}), and AT2019ntm (detection of a potential H$\alpha$ emission line corresponding to $z=0.116$, outside the LVC $2\sigma$ confidence region of GW190814; \citealt{ackley20}). See also Table~\ref{tab:tns} (Appendix~\ref{app:TNS}) for MegaCam photometry of these sources.

Wide-field imaging across multiple wavelengths allowed for coverage of the entire localization region, and many teams were able to acquire these images within $\lesssim1$ day post-merger\footnote{The coverage for this event can be visualized and quantified on the \href{http://treasuremap.space/alerts?graceids=S190814bv}{GW TreasureMap} (\citealt{wyatt20}). We have submitted our CFHT MegaCam pointings to the TreasureMap and we encourage all who engage in EM follow-up of GWs to do so for future events.}. However, to date, no compelling EM counterpart has been found for GW190814. Some candidates in Figure~\ref{fig:reportedtransients} were never conclusively disqualified or even classified, likely because they were far too bright to correspond to the kilonova signatures predicted by various models (e.g. \citealt{kawaguchi16, fernandez17, tanaka18}). As discussed in Section~\ref{ssc:obs}, a depth of $i\sim22.0$ was likely required to observe a kilonova at the location of GW170817. A detailed analysis of GW190814 in particular (\citealt{kawaguchi20}) similarly suggests that observations deeper than $i,~z \sim 22.0$ within 2 days post-merger were required to detect a kilonova counterpart to the event. The mean depths achieved by surveys such as the Zwicky Transient Facility (ZTF; $r,~i\sim$ 20.0 at 2 days; \citealt{singer19}) were therefore insufficient for this purpose. Furthermore, while facilities such as Pan-STARRS played an essential role in disqualifying several candidates and guiding follow-up of compelling candidates, the depth reached by Pan-STARRS ($z~\sim21.9$ at 1.5 days; \citealt{ackley20}) was also insufficient to search for kilonovae.  

The observing campaigns carried out with DECam and the CFHT MegaCam program described here have the deepest coverage yet reported. This underscores the need for large-aperture telescopes with wide-field imagers in following up mergers as distant as GW190814. We consider the following $5\sigma$ CFHT MegaCam limiting magnitudes (shown in Figure~\ref{fig:depth}) in the remainder of our analysis:
\begin{itemize}
    \item $g > 22.8,$ 1.7 days
    \vspace{-6pt}
    \item $i > 23.1,$ 3.7 days
    \vspace{-6pt}
    \item $i > 22.8,$ 4.7 days
    \vspace{-6pt}
    \item $i > 23.9,$ 8.7 days
\end{itemize}

The mean areal coverage of the 50\% localization region during these observations is 93.1\% and mean areal coverage of the 50\% $< p <$~90\% localization region is 34.3\%. The total integrated probability coverage ranges from 61.5\% to 70.5\% for these observations. We demonstrate the value of our MegaCam observations in constraining the parameters of the presumed NS-BH merger kilonova in the following section.

\section{Constraints on a Possible Kilonova}\label{sec:models}

Whether or not a kilonova is produced during a NS-BH merger is highly sensitive to the parameters of the initial binary. To produce a kilonova, the NS must be tidally disrupted outside of the BH's innermost stable circular orbit (ISCO). This outcome is most likely in systems involving small binary mass ratios $q = \frac{M_{\mathrm{BH}}}{M_{\mathrm{NS}}}$ (e.g. $M_{\mathrm{BH}}\lesssim8~M_\odot$, \citealt{shibata08, lovelace13, foucart14, foucart18, foucart19}) and/or highly-spinning black holes (e.g. $\chi_{\mathrm{BH}}\gtrsim0.7$, \citealt{etienne09, lovelace13, foucart14, kawaguchi15}). Furthermore, the mass, radius, and equation of state of the progenitor NS can also impact the mass of the dynamical ejecta and the remnant BH's accretion disk produced by this tidal disruption (\citealt{shibata08, kawaguchi15}). In particular, a sufficiently compact NS could avoid this disruption completely until it is beyond the BH's ISCO and thus produce no EM signature. Finally, the orientation of magnetic fields around the merger remnant can also impact mass outflows (e.g. \citealt{barnes16, christie19}). 

Assuming that a kilonova does occur, the spectral and temporal behaviour are sensitive to the mass $M_{\mathrm{ej}}$, velocity $v_{\mathrm{ej}}$, and opacity $\kappa_{\mathrm{ej}}$ of the ejecta. Below, we use a simple kilonova model, which is largely progenitor agnostic (NS-NS or NS-BH), to constrain $M_{\mathrm{ej}}$, $v_{\mathrm{ej}}$, and $\kappa_{\mathrm{ej}}$ using the MegaCam limits shown in Figure~\ref{fig:depth} and listed in Section~\ref{sec:other-searches}.

\subsection{Kilonova Model}\label{ssc:knovamodel}
We use a simple 1D kilonova model based on that described in \cite{metzger19}. This model assumes a centrally-concentrated energy source and homologous expansion of a surrounding single-zone ejecta powered by the radioactive decay of heavy $r$-process elements. We use a radioactive heating rate fitting formula (\citealt{korobkin12}): 
\begin{equation}\label{eq:heatingrate}L_{\mathrm{in}}(t) = C\cdot M_{\mathrm{ej}}\left(0.5-\pi^{-1}\arctan{\left(\frac{t-t_0}{\sigma}\right)}\right)^{1.3},
\end{equation}
where $C = 4\times10^{18}~\mathrm{cm^{2}~s^{-3}}$, $t_0 = 1.3 ~\mathrm{s}$, and $\sigma = 0.11~\mathrm{s}$ are constants, and $t$ is the time post-merger in seconds. No other energy sources are considered. Equation~(\ref{eq:heatingrate}) is most accurate for ejecta composed largely of lanthanides and/or actinides, i.e., the heaviest $r$-process elements (\citealt{metzger19}). This assumption is not necessarily valid for NS-NS mergers, in which a short- or long-lived merger remnant such as a hyper-massive NS can produce a large neutrino flux, raise the electron fraction $Y_{e}$ of the surrounding ejecta, lower the number of free neutrons, and inhibit production of these elements via the channel $\nu_{e} + n \rightarrow p + e^{-}$ (\citealt{lippuner17}). Crudely, $Y_{e}\lesssim0.25$ corresponds to lanthanide-rich ejecta and will produce a `red' kilonova including a significant heating contribution from the decay of these lanthanides, whereas $Y_{e}\gtrsim0.25$ corresponds to lanthanide-poor ejecta and will produce a `blue' kilonova powered by the decay of $r$-process elements lighter than the lanthanides (e.g. \citealt{metzger19}, their Figure 6). In the case of a NS-BH merger, the remnant must be a BH (e.g. \citealt{metzger19}, their Figure 18), so we may in general expect a lower neutrino flux, a smaller $Y_{e}$, and a redder kilonova with some contribution from lanthanides. 

We also define a time-dependent thermalization efficiency, as in \cite{barnes16}:
\begin{equation}\label{eqn:therm}
\epsilon_{\mathrm{th}}(t) = 0.36\left[e^{-at_{\mathrm{days}}} + \frac{\ln({1+2b t_{\mathrm{days}}^{d}})}{2b t_{\mathrm{days}}^{d}}\right],
\end{equation}
where $t_{\mathrm{days}}$ is the time in days, and $a, b, d$ are fitting parameters of order unity which depend on $M_{\mathrm{ej}}$ and $v_{\mathrm{ej}}$. These parameters are tabulated in Table 1 of \cite{barnes16}, and are fit assuming a randomly-oriented magnetic field in the ejecta. We linearly interpolate their parameters to obtain fitting parameters at other masses and velocities. With the radioactive heating rate defined in Equation~(\ref{eq:heatingrate}) and thermalization efficiency defined in Equation~(\ref{eqn:therm}), we then describe the total bolometric luminosity as in \cite{chatzopoulos12}:

\begin{equation}\label{eqn:chatz}
L_{\mathrm{bol}}(t) = \frac{2}{t_d^2}\exp \left({\frac{-t^2}{t_d^2}}\right)\int_0^t L_{\mathrm{in}}(t') \epsilon_{\mathrm{th}}(t') \exp \left({\frac{t'^2}{t_d^2}}\right) t' dt',
\end{equation}
where $t_d =  \sqrt{2\kappa_{\mathrm{ej}} M_{\mathrm{ej}}/\beta v_{\mathrm{ej}} c}$ is a diffusion timescale of the system, $\kappa_{\mathrm{ej}}$ is the grey (frequency-independent) opacity, and $\beta=13.8$ is a parameter based on the geometry of the system. In Equation (\ref{eqn:chatz}), we have assumed that the original photosphere radius is vanishingly small and we have neglected the initial thermal energy of the system. We assume that the expanding material is well-described by a blackbody. This material rapidly cools and the radius of the photosphere expands until it reaches a critical temperature floor $T_c\approx2500~\mathrm{K}$ near the recombination temperature of lanthanides, at which point the temperature becomes fixed and the photosphere begins to recede towards the central engine (\citealt{barnes13}). The temperature and radius of the photosphere are completely determined by $L_{\mathrm{bol}}$ at a given point in time:
\begin{equation}
T_{\mathrm{phot}}(t) = \max\left[ \left( \frac{L_{\mathrm{bol}}(t)}{4\pi\sigma_{\mathrm{sb}}v_{\mathrm{ej}}^2 t^2}\right)^{1/4},~T_c     \right],
\end{equation}
\[R_{\mathrm{phot}}(t) =
\begin{cases}
 v_{\mathrm{ej}} t \quad& \text{if}~T_{\mathrm{phot}} > T_c\\
 \left( \frac{L_{\mathrm{bol}}(t)}{4\pi\sigma_{\mathrm{sb}}T_c^4}\right)^{1/2} & \text{if}~T_{\mathrm{phot}} \leqslant T_c
\end{cases},\]
where $\sigma_{\mathrm{sb}}$ is the Stefan-Boltzmann constant, allowing us to construct the evolving thermal spectral energy distribution (SED) of the kilonova and produce $g$- and $i$-band magnitude predictions for comparison to the limits achieved by MegaCam.

\begin{figure*} [!ht]
\center{
\includegraphics[scale=0.31]{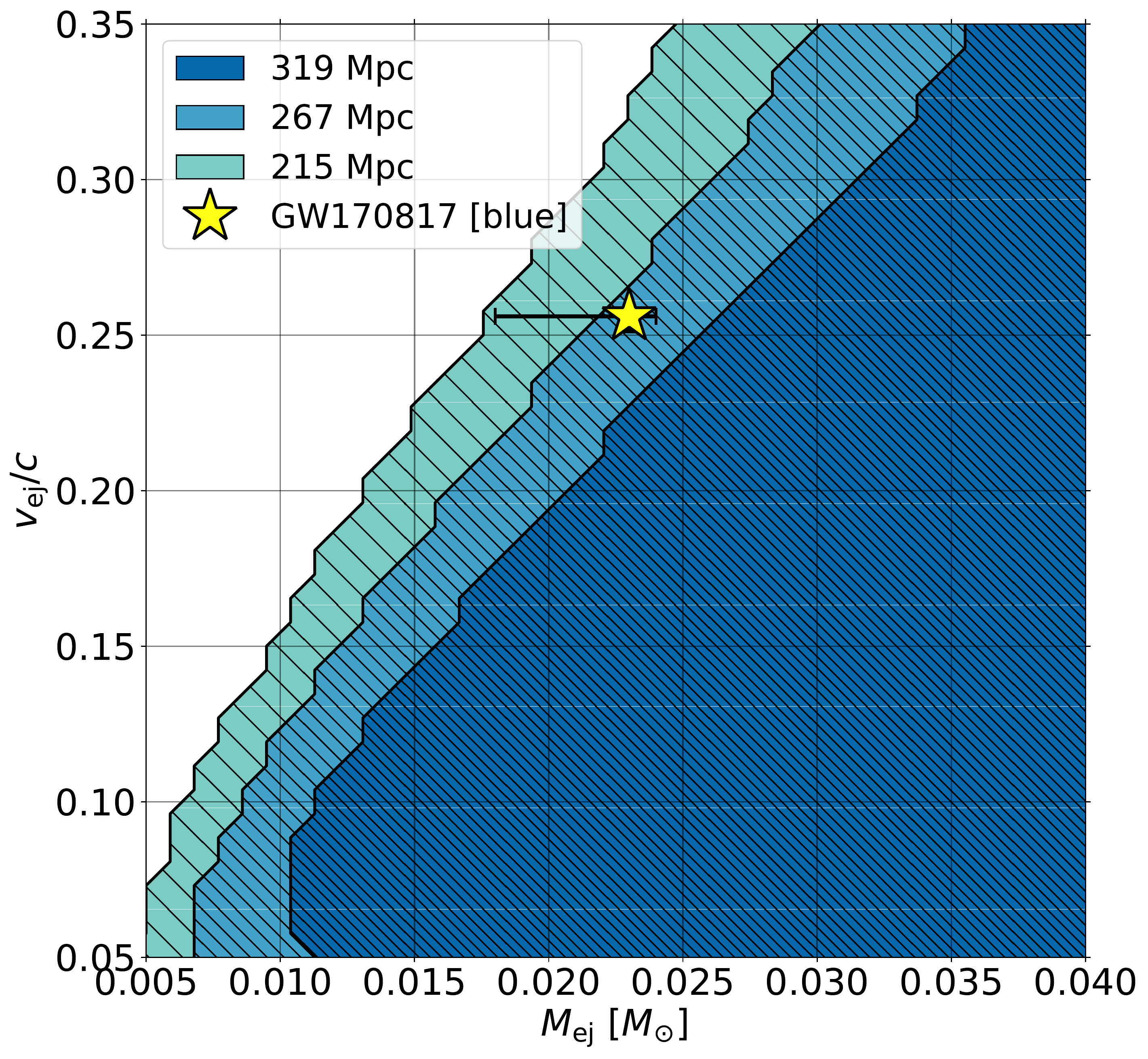}
\includegraphics[scale=0.31]{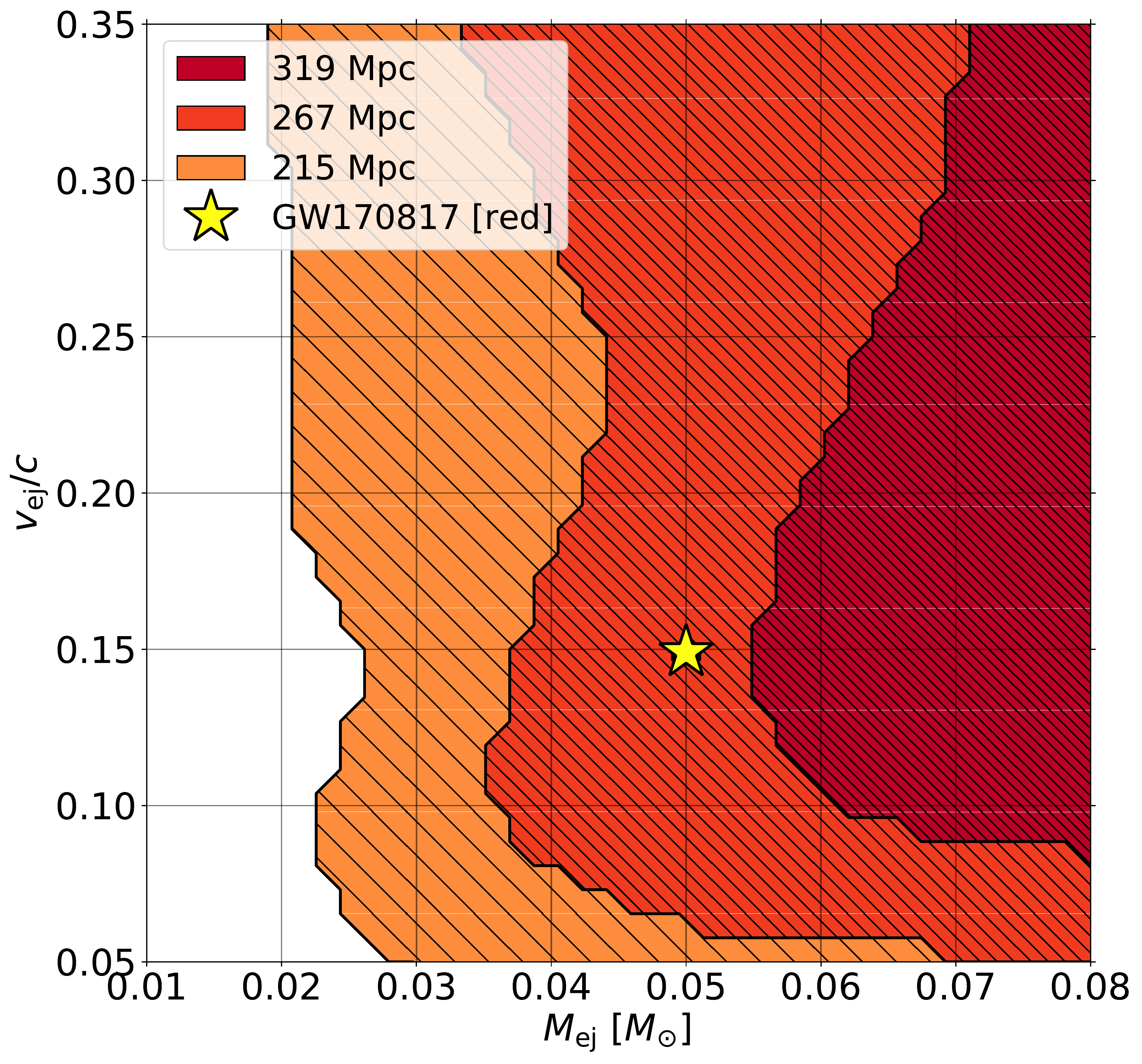}
}
\figcaption{\textbf{Allowed ejecta masses and velocities of the kilonova, constrained using multi-band limits imposed by MegaCam.} Colored regions denote parameters which are ruled out using the model outlined in Section~\ref{ssc:knovamodel}. In general, assuming a smaller distance to GW190814 places more strict limits on the parameters of the ejecta.  In the left panel, a constant opacity $\kappa_{\mathrm{ej}}=0.5\mathrm{~cm^2 g^{-1}}$ is assumed. The best-fit parameters and associated uncertainties for the blue component of a 2-component kilonova model for GW170817 (\citealt{villar17}) are presented as well (yellow star). In the right panel, a constant opacity $\kappa_{\mathrm{ej}}=5\mathrm{~cm^2 g^{-1}}$ is assumed, and the best-fit parameters for the red component of the same 2-component kilonova model for GW170817 (\citealt{villar17}) are shown (yellow star). Uncertainties on the parameters of this red component are on the order of point size. Note that the right panel explores a different range of $M_{\mathrm{ej}}$ from the left panel, since larger $M_{\mathrm{ej}}$ are conclusively ruled out in the left panel.}
\label{fig:mej_vej_space}
\end{figure*}

\subsection{MegaCam Constraints on Kilonova Parameters}\label{ssc:megacam_constraints}
We explore the allowed ejecta masses and velocities for fixed grey opacities of $\kappa_{\mathrm{ej}}=0.5\mathrm{~cm^2 g^{-1}}$ and $\kappa_{\mathrm{ej}}=5\mathrm{~cm^2 g^{-1}}$ in Figure~\ref{fig:mej_vej_space}. Colored regions show the parts of parameter space which are ruled out. A region is ruled out if the $g$- and/or $i$-band light curve predicted by our model is inconsistent with the limits imposed by MegaCam (Figure~\ref{fig:depth} and Section~\ref{sec:other-searches}). For each ($M_{\mathrm{ej}}$, $v_{\mathrm{ej}}$), we consider distances of 215 Mpc, 267 Mpc, and 319 Mpc to the source, representing the uncertainty in the luminosity distance $d_L = 267 \pm 52$ Mpc of GW190814. In all cases, assuming a smaller distance to the source imposes tighter limits on the parameters of the ejecta, and on $M_{\mathrm{ej}}$ in particular. For a blue kilonova with $\kappa_{\mathrm{ej}}=0.5\mathrm{~cm^2 g^{-1}}$, the constraints depend strongly on both the ejecta mass and velocity. If we select a fiducial ejecta velocity $v_{\mathrm{ej}}=0.2c$ (typical velocity of tidal tails and/or disk outflows, e.g. \citealt{fernandez17, christie19, metzger19}), we are able to impose the constraint that $M_{\mathrm{ej}} \lesssim 0.015M_{\odot}$. For a progenitor NS of mass $1.4M_{\odot}$, this corresponds to $\lesssim1$\% of the mass being ejected. For a more lanthanide-rich, red kilonova ($\kappa_{\mathrm{ej}}=5\mathrm{~cm^2 g^{-1}}$), there is less structure in the constrained velocities in parameter space. We are therefore unable to place any meaningful constraints on the velocity of the ejecta for a red kilonova. Nonetheless, we are able to impose the constraint that $M_{\mathrm{ej}} \lesssim 0.04M_{\odot}$ for such a red kilonova for arbitrary ejecta velocity. This translates to $\lesssim3$\% mass ejection for a progenitor NS of mass $1.4M_{\odot}$. Figure~\ref{fig:mej_vej_space} also shows the best-fit parameters for each individual component of the 2-component symmetric kilonova model of GW170817 presented in \cite{villar17}. The `blue' ($\kappa_{\mathrm{ej}}^{\mathrm{blue}}=0.5\mathrm{~cm^2 g^{-1}}$; fixed) component of this model, with $M_{\mathrm{ej}}^{\mathrm{blue}}=0.023^{+0.005}_{-0.001}M_{\odot}$ and $v_{\mathrm{ej}}^{\mathrm{blue}}=0.256^{+0.005}_{-0.002}c$, is compared to our $\kappa_{\mathrm{ej}}=0.5\mathrm{~cm^2 g^{-1}}$ constraints. The `red' ($\kappa_{\mathrm{ej}}^{\mathrm{red}}=3.65^{+0.09}_{-0.28}\mathrm{~cm^2 g^{-1}}$; fit parameter) component of this model, with $M_{\mathrm{ej}}^{\mathrm{red}}=0.050^{+0.001}_{-0.001}M_{\odot}$ and $v_{\mathrm{ej}}^{\mathrm{red}}=0.149^{+0.001}_{-0.002}c$, is compared to our $\kappa_{\mathrm{ej}}=5\mathrm{~cm^2 g^{-1}}$ constraints. Note that this 2-component kilonova model is not the same as the model shown in Figure~\ref{fig:depth}. As we are using a 1-component model and comparing to individual components of a 2-component model, our MegaCam observations rule out a 1-component kilonova with parameters similar to either the red or the blue component of the GW170817 kilonova, and not the GW170817 kilonova altogether. 

\begin{figure}[t!]
\center{
\includegraphics[scale=0.30,angle=0]{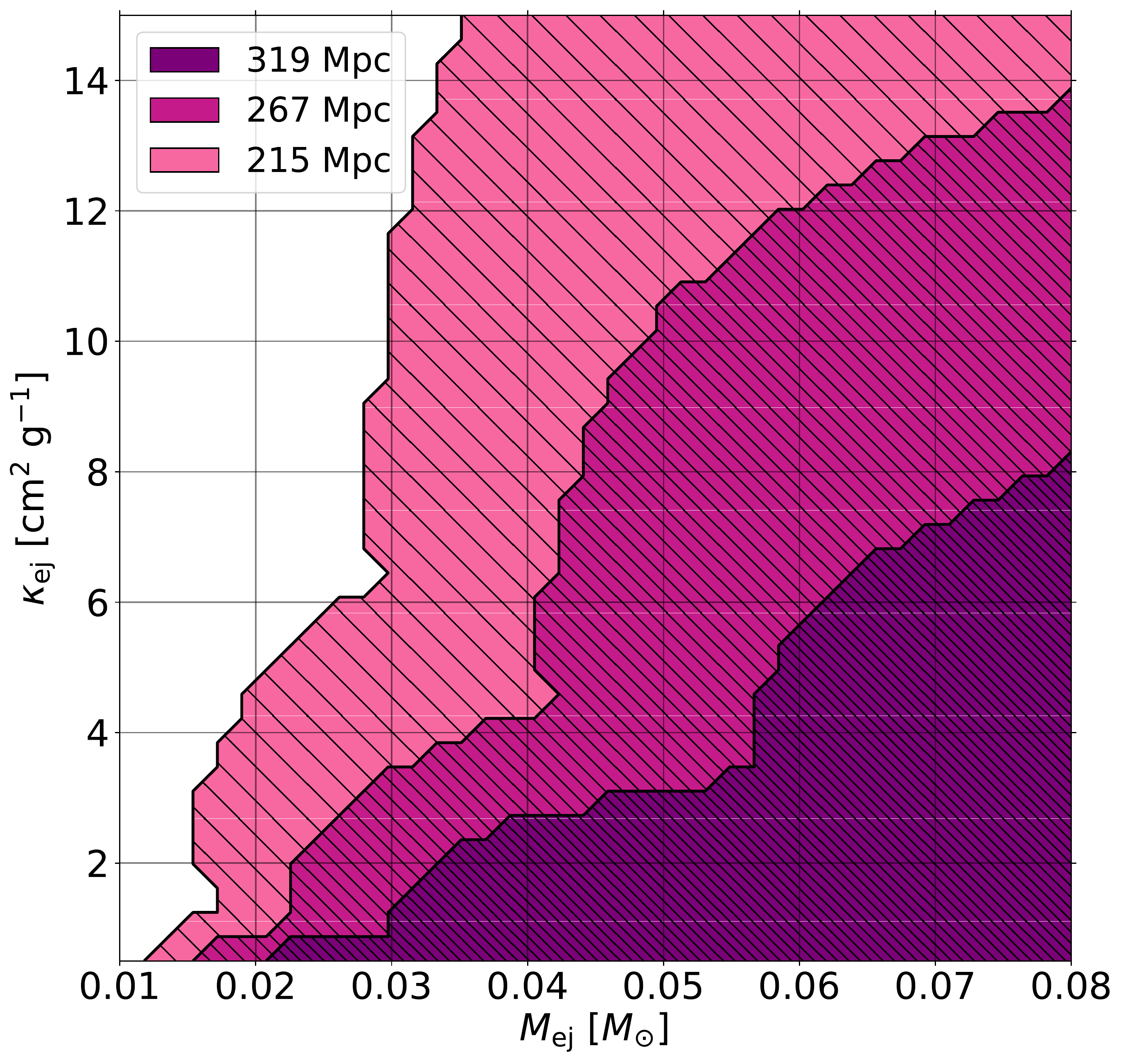}
}
\figcaption{\textbf{Allowed ejecta masses and opacities of the kilonova for a fixed $\mathbf{v_{\mathrm{ej}}=0.2c}$}. Colored regions denote parameters which are ruled out. Our constraint on $M_{\mathrm{ej}}\gtrsim0.04M_{\odot}$ is essentially the same for opacities in the range $\kappa_{\mathrm{ej}}=5 - 10\mathrm{~cm^2 g^{-1}}$.}
\label{fig:mej_kej_space}
\end{figure}

We also explore the ($M_{\mathrm{ej}}, \kappa_{\mathrm{ej}}$) parameter space for a fixed $v_{\mathrm{ej}}=0.2c$ in Figure~\ref{fig:mej_kej_space}. Masses are most constrained for ejecta with low opacities. For a lanthanide-rich merger ejecta with opacity $\kappa_{\mathrm{ej}}=5 - 10\mathrm{~cm^2 g^{-1}}$, our constraint $M_{\mathrm{ej}} \lesssim 0.04M_{\odot}$ is essentially the same as that presented in Figure~\ref{fig:mej_vej_space}. Since our constraints for a red kilonova are more conservative than for a blue one, and since we may in general expect a redder kilonova from a NS-BH merger, we focus on these constraints. In all, for a kilonova with opacity $\kappa_{\mathrm{ej}}=5 - 10\mathrm{~cm^2 g^{-1}}$ at the mean distance of GW190814, we impose an upper limit on the ejecta mass $M_{\mathrm{ej}} \lesssim 0.04M_{\odot}$. We also rule out a 1-component red or blue kilonova with parameters similar to the corresponding individual red or blue components of GW170817. In order to compute a confidence level for these constraints, we compare the model predictions to the limiting magnitudes of each field individually (see Table~\ref{tab:details} in Appendix~\ref{app:obs}), record which fields are able to constrain $M_{\mathrm{ej}} \lesssim 0.04M_{\odot}$, and compute the total integrated probability of the localization region which is covered by the fields which apply this constraint. At 4.7 days and 8.7 days, none of the observations are able to constrain $M_{\mathrm{ej}} \lesssim 0.04M_{\odot}$. At 3.7 days, however, the total integrated probability of the fields which successfully apply this constraint is 61.5\%, and so we impose this constraint at 61.5\% confidence. The inability of the images at 4.7 days to provide any additional constraints is not surprising, given that these images were shallower than those of the previous night, whereas the kilonova is expected to fade by as much as $\sim$0.5 mag over this period. Similarly, the significant fading by 8.7 days makes it difficult to derive any constraints at these late epochs. The constraint $M_{\mathrm{ej}} \lesssim 0.04M_{\odot}$ is in agreement with those of other teams ($M_{\mathrm{ej}} \lesssim 0.05M_{\odot}$, \citealt{andreoni20}; $M_{\mathrm{ej}} \lesssim 0.1M_{\odot}$, \citealt{ackley20}), although we note that \cite{andreoni20} propose a less lanthanide-rich ejecta with opacity $\kappa_{\mathrm{ej}}<2\mathrm{~cm^2 g^{-1}}$ for their limit. 

For the case of a low-opacity blue kilonova ($\kappa_{\mathrm{ej}}=0.5\mathrm{~cm^2 g^{-1}}$), the MegaCam limit $g > 22.8$ at 1.7 days is the most constraining and is necessary to rule out the smallest ($M_{\mathrm{ej}} \lesssim 0.02M_{\odot}$) masses in parameter space for small ejecta velocities. For a more lanthanide-rich red kilonova ($\kappa_{\mathrm{ej}}=5 - 10\mathrm{~cm^2 g^{-1}}$), the most constraining MegaCam limits are those at 3.7 days. Limits from other epochs do not provide any additional constraints at these large opacities. This is in agreement with the observations of \cite{andreoni20}, who similarly find their DECam limit $z > 22.3$ at 3.4 days to be the most constraining. \cite{kawaguchi20} also find this limit to be the most constraining among the DECam limits in their analysis. These observations can inform observing strategies in future follow-up campaigns of NS-BH and NS-NS mergers with UV/optical/near-IR instruments. To enable detection of these events, future campaigns should focus on maximizing the depth achieved by their images, and especially on early ($\lesssim1$ day) imaging in the UV/optical (e.g. $u$-, $g$-band) followed immediately by a transition to the $i$- and/or $z$-bands. Obtaining these deep early images may be more effective for identifying a counterpart than extended observing campaigns in these bands.   

\section{Conclusions}\label{sec:conclusions}
We describe a series of deep, wide-field, multi-band CFHT MegaCam observations of the localization region of the LIGO/Virgo GW signal GW190814/S190814bv, which represents the first robust detection of a possible NS-BH merger. We employ a hybrid observing strategy of wide-field tiling of the 50\% GW localization region, and galaxy-targeted observations in the 50\% $< p <$~90\% localization region, from 1.7 days to 8.7 days post-merger, in search of an EM counterpart to GW190814. We use image differencing and a convolutional neural network to detect and classify candidate transients for further assessment as potential counterparts to the GW merger. We find no convincing EM counterparts to GW190814 in our images. This suggests that either (1) the NS in the binary was not tidally disrupted outside of the BH's ISCO prior to the merger, (2) the transient lies outside of the observed sky footprint, or (3) the lighter object is a low-mass BH. We therefore apply our measured $5\sigma$ depth of $g > 22.8$ (AB mag) at 1.7 days post-merger and depths of $i > 23.1$ and $i > 23.9$ at 3.7 and 8.7 days post-merger, respectively, to constrain the parameters of the presumed kilonova which accompanied the merger. The total integrated probability of these MegaCam observations ranges from 61.5\% to 70.5\%. Using a simple 1D single-component, single-zone kilonova model and the MegaCam limit $i > 23.1$ at 3.7 days, we are able to constrain the ejecta mass to $M_{\mathrm{ej}} \lesssim 0.04M_{\odot}$ ($\lesssim3$\% mass ejection for a progenitor NS of mass $1.4M_{\odot}$) for a lanthanide-rich merger ejecta with $\kappa_{\mathrm{ej}}=5 - 10\mathrm{~cm^2 g^{-1}}$ at the mean LIGO/Virgo luminosity distance of 267 Mpc to GW190814 at 61.5\% confidence. We also set the limit $M_{\mathrm{ej}} \lesssim 0.015M_{\odot}$ ($\lesssim1$\% mass ejection for a progenitor NS of mass $1.4M_{\odot}$) for a lanthanide-poor merger ejecta with $\kappa_{\mathrm{ej}}=0.5\mathrm{~cm^2 g^{-1}}$, using the MegaCam limit $g > 22.8$ at 1.7 days, at 65.5\% confidence.

The limits imposed by our observing campaign are among the most strict optical/near-IR limits for this source. Should the LVC confirm that the mass of the lighter object is within the realm of possibility for a NS, these limits will be valuable in further constraining the system. Regardless of the eventually reported masses, our observations reiterate the importance of maximizing depth in high-cadence UV/optical/IR imaging of these events. The current detection horizons for NS-NS and NS-BH mergers in O3 for Advanced LIGO are 110-130 Mpc and 190-240 Mpc, respectively, and are expected to extend to 160-190 Mpc and 300-330 Mpc for Advanced LIGO in the fourth LIGO/Virgo observing run, O4 (\citealt{abbottLIGO18}). Follow-up with 3 m and larger aperture telescopes with wide-field imaging instruments will be crucial to detecting EM counterparts to these mergers. Observatories such as the CFHT and instruments like MegaCam can play an important role in the first detection of an EM counterpart to a NS-BH merger and our transient detection pipeline described here can aid in realizing this goal.

\acknowledgments
We thank the CFHT queued service observing team and the telescope staff for their help in obtaining these observations. We also thank the anonymous referee for their helpful feedback, which has strengthened this study. N.V. acknowledges funding from the Natural Sciences and Engineering Research Council of Canada (NSERC), Fonds de recherche du Qu\'ebec - Nature et Technologies (FRQNT), and the Bob Wares Science Innovation Prospectors Fund. J.J.R.\ and D.H.\ acknowledge support from a NSERC Discovery Grant, a  FRQNT Nouveaux Chercheurs Grant, and support from the Canadian Institute for Advanced Research (CIFAR).  J.J.R.\ acknowledges funding from the McGill Trottier Chair in Astrophysics and Cosmology, the McGill Space Institute, and the Dan David Foundation.
\newline
\facility{CFHT}
\software{\href{https://docs.astropy.org/en/stable/}{\texttt{astropy}} (\citealt{astropy18}), \href{https://photutils.readthedocs.io/en/stable/}{\texttt{photutils}} (\citealt{photutils19}), \href{https://github.com/cds-astro/mocpy}{\texttt{MOCpy}} (\citealt{fernique14}), \href{https://lscsoft.docs.ligo.org/ligo.skymap/}{\texttt{ligo.skymap}}, \href{https://github.com/guillochon/MOSFiT}{\texttt{MOSFiT}} (\citealt{guillochon18}) \href{https://github.com/toros-astro/astroalign}{\texttt{astroalign}} (\citealt{beroiz19}), \href{https://github.com/acbecker/hotpants}{High Order Transform of PSF And Template Subtraction} (\citealt{becker15}), \href{https://github.com/dmitryduev/braai}{\texttt{braai}} (\citealt{duev19}), \href{https://www.tensorflow.org/}{\texttt{TensorFlow}}, \href{https://scikit-learn.org/stable/index.html}{\texttt{scikit-learn}}}

\bibliographystyle{apj}

\appendix{}
\section{Real-bogus training results}\label{app:rb}
\begin{figure} [!ht]
\center{
\includegraphics[scale=0.36,angle=0]{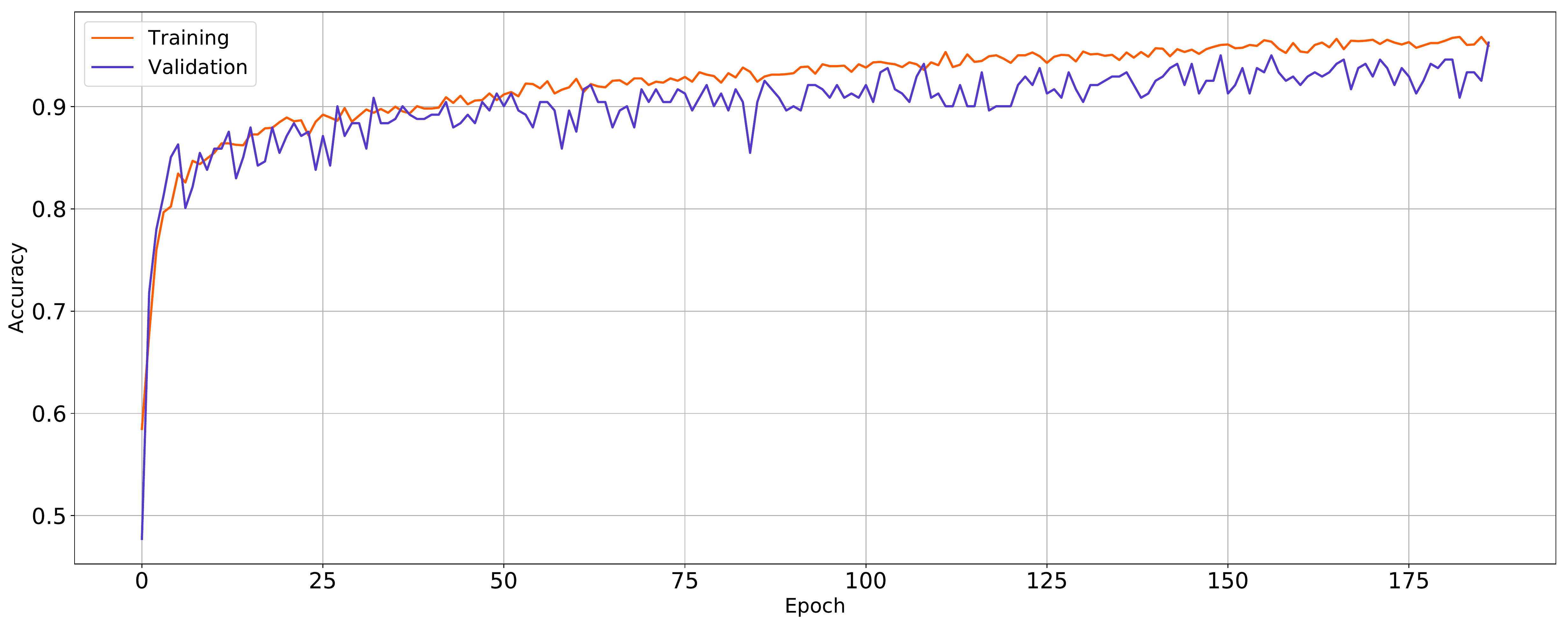}
}
\figcaption{\textbf{Training/validation accuracy versus epoch (iteration) of training.} Training is enabled for a maximum of 200 epochs. Early stoppage is employed to prevent the model from over-fitting to the training set by forcing the training to stop if the validation accuracy does not show any improvement over 50 epochs. In practice, the training set and validation set converge to $\sim$95\% accuracy and plateau in 160-190 epochs. In this training, early stoppage was activated at 187 epochs.
}\label{fig:training_accuracy}
\end{figure}

\begin{figure*} [!t]
    \centering
    \includegraphics[scale=0.6,angle=0]{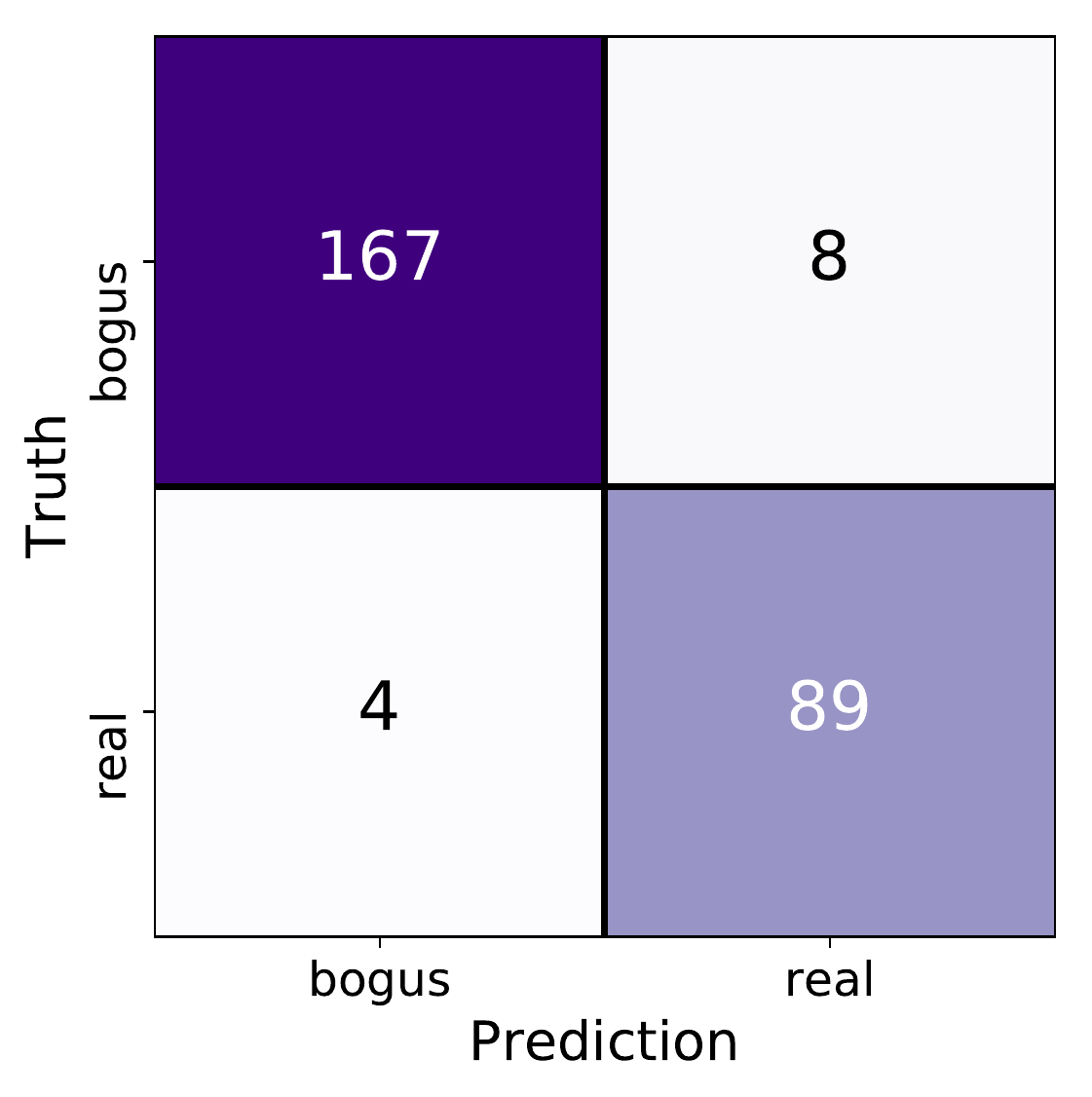}
    \includegraphics[scale=0.6,angle=0]{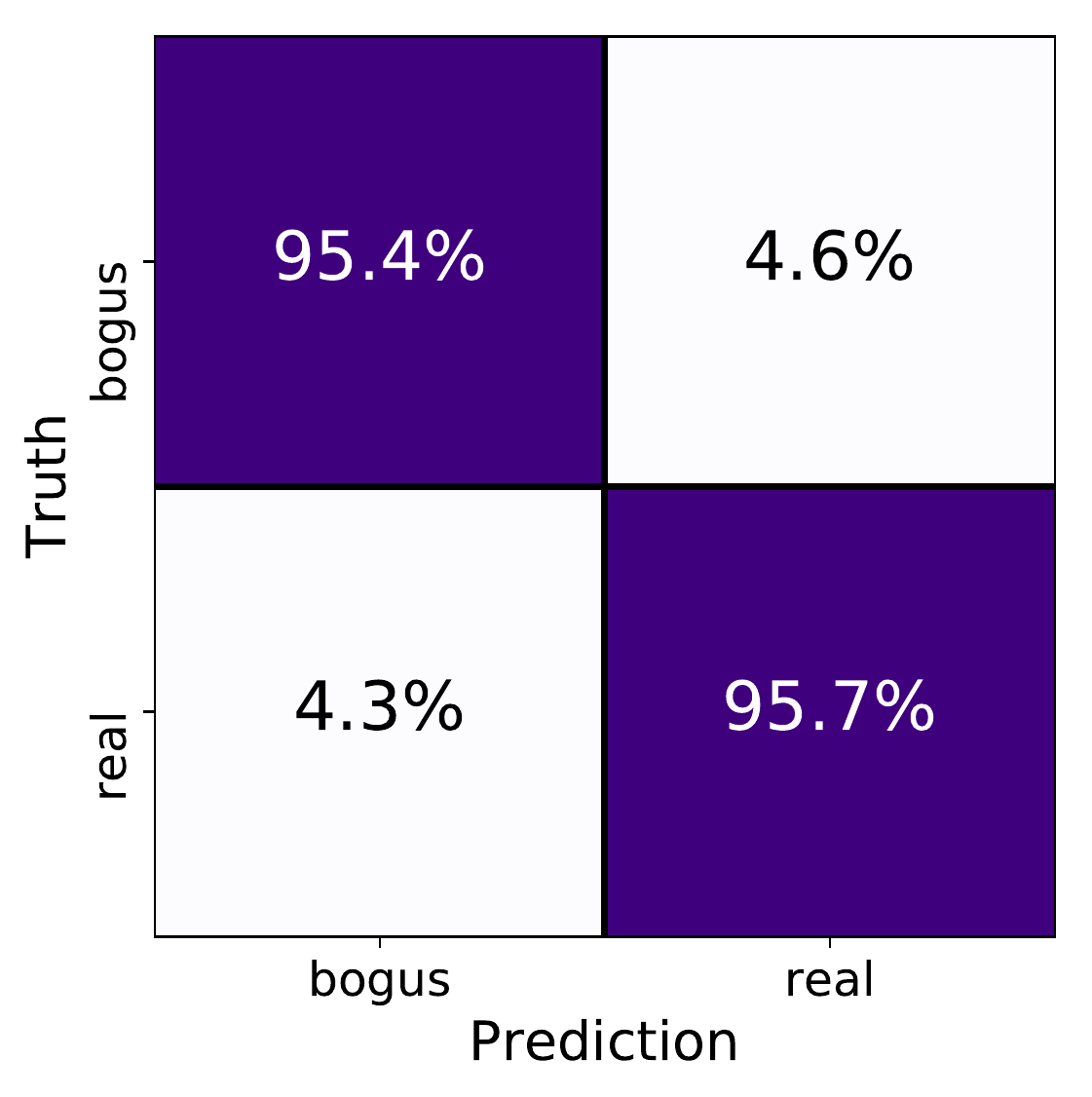}
    \figcaption{\textbf{Non-normalized (left) and normalized (right) confusion matrices of the model predictions for the 268 sources in the test set, for Real-Bogus score threshold RB~$\mathbf{\geqslant0.5}$.} The normalized matrix shows the true/false positive/negative rates. For this model, if a Real-Bogus score of RB $\geqslant$ 0.5 is used to define a real source, then the false positive rate (FPR) is 4.6\% and the false negative rate (FNR) is 4.3\%.}\label{fig:confusion_matrix}
\end{figure*}

\begin{figure} [!ht]
\center{
\includegraphics[scale=0.64,angle=0]{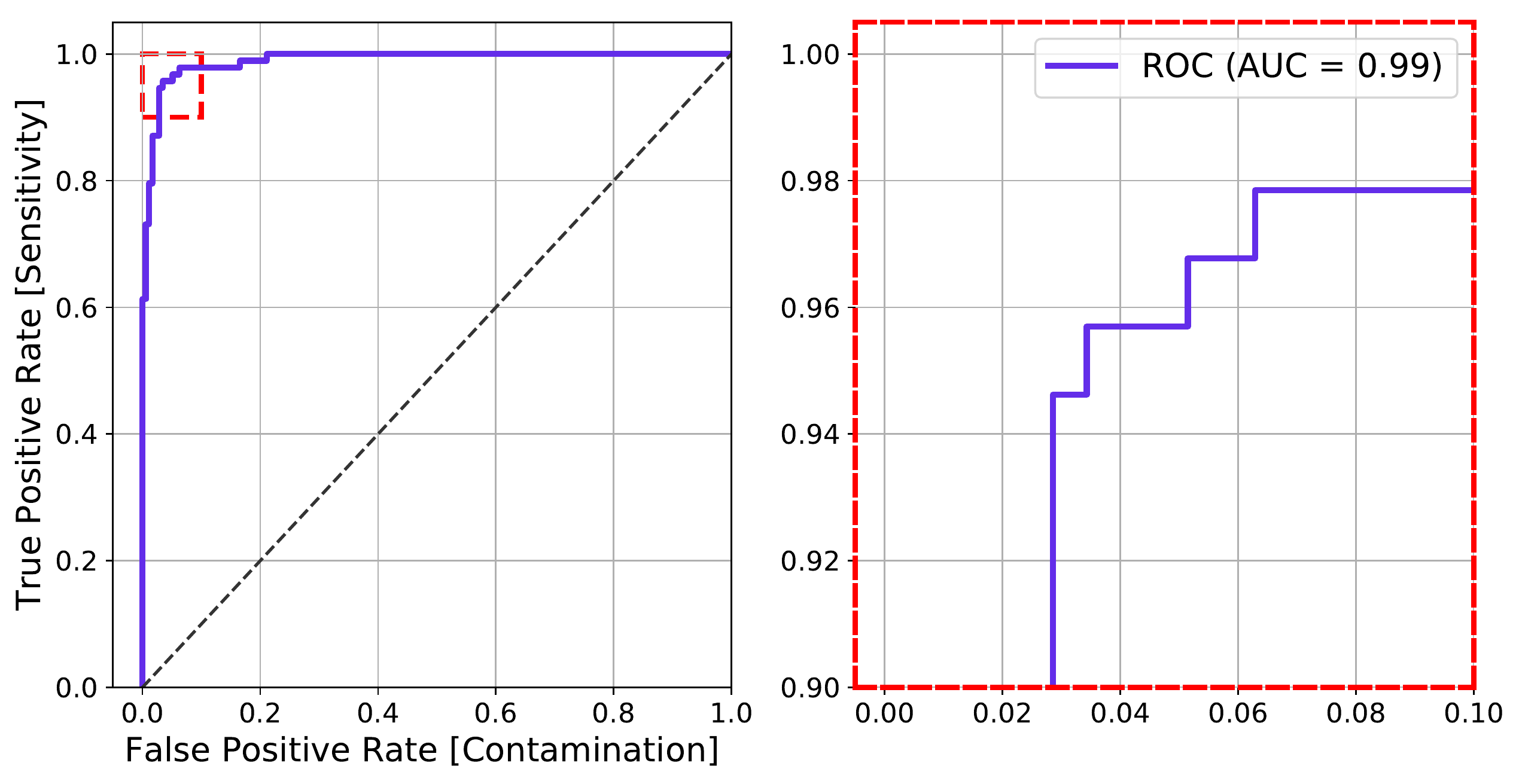}
}
\figcaption{\textbf{The receiver operating characteristic (ROC) curve for the model.} An inset is shown in the right panel to highlight the section of the curve with the most structure. RB $\geqslant$ 0.5 denotes a real source. The area under the ROC curve (AUC) is a measure of how informative the model is. The dotted line ($\mathrm{AUC}=0.5$) denotes a completely random classifier. For $\mathrm{AUC}\approx0.99$, given a randomly-selected true real source and a randomly-selected true bogus source, the classifier has a 99\% probability to favour the true real as being real over the true bogus.}\label{fig:ROC}
\end{figure}

\begin{figure} [!ht]
\center{
\includegraphics[scale=0.57,angle=0]{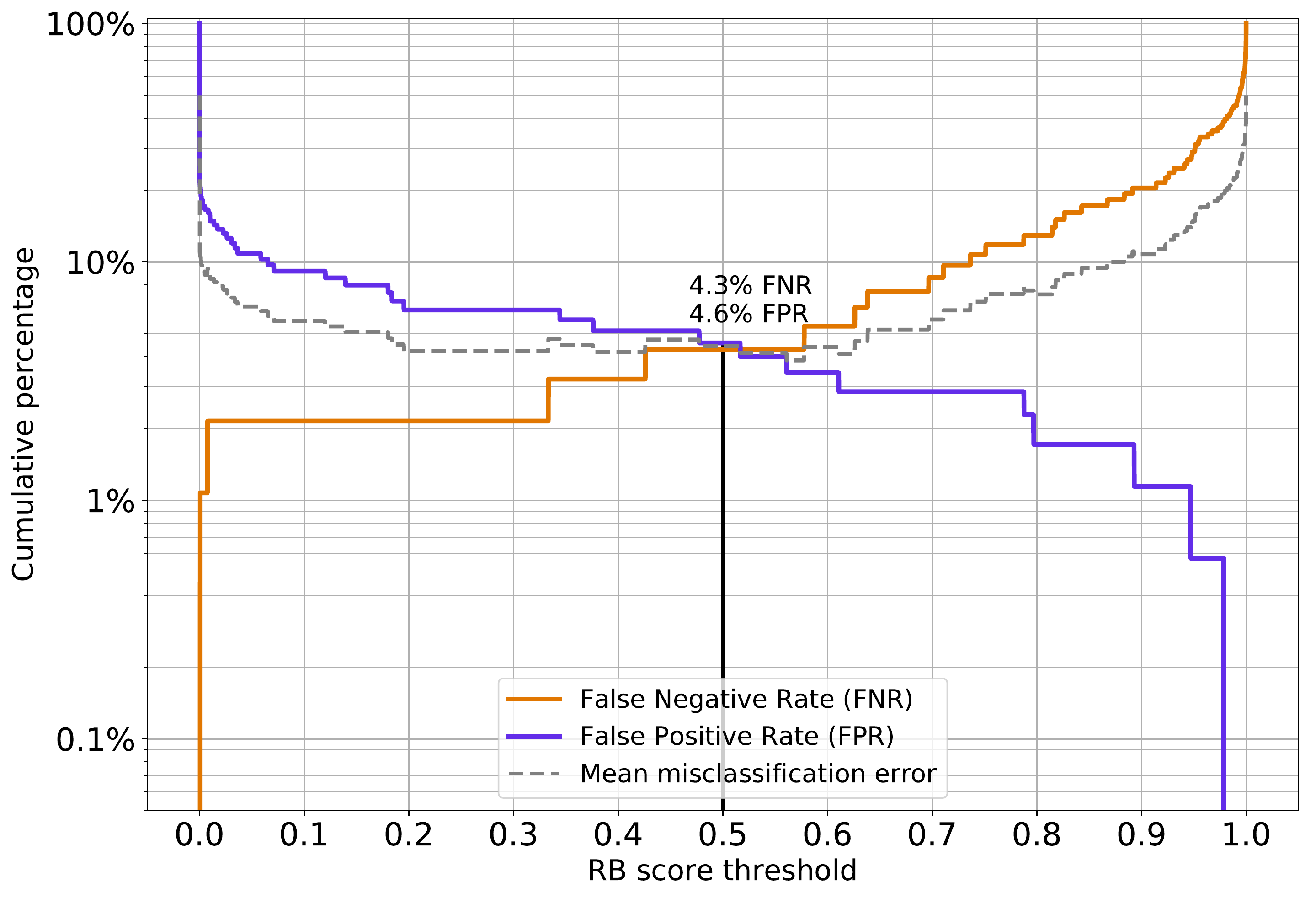}
}
\figcaption{\textbf{The measured false positive rate (FPR) and false negative rate (FNR) as a function of Real-Bogus (RB) score threshold.} These rates were computed using the 268 sources in the test set. For an RB score threshold of $\mathrm{RB}\geqslant0.5$, FPR = 4.6\% and FNR = 4.3\%. The mean misclassification error is therefore 4.5\% at this threshold.}\label{fig:FPR_FNR}
\end{figure}

Here, we present the results of training the \texttt{braai} neural network on the dataset described in Section~\ref{ssc:braai}. 
Figure~\ref{fig:training_accuracy} shows the evolution of the model's classification accuracy for the training set and validation set at each epoch (iteration) of training. For imbalanced datasets, the accuracy alone is not a reliable indicator of the usefulness of a model. Considering the example of a dataset which is 95\% bogus, 5\% real, a classifier that flags all sources as bogus would have 95\% accuracy, but would not be useful for detecting transient sources. We therefore also compute the confusion matrices (Figure~\ref{fig:confusion_matrix}) and receiver operating characteristic (ROC) curve (Figure~\ref{fig:ROC}) resulting from applying the trained model to the 268 sources in the test set. The confusion matrices succinctly present the true positive rate, false positive rate, true negative rate, and false negative rate (TPR, FPR, TNR, FNR). The ROC curve shows the sensitivity (i.e. TPR) of the model as a function of contamination (i.e. FPR). Both confusion matrices and the ROC curve are computed for a Real-Bogus score of RB $\geqslant$ 0.5 denoting a real source. Finally, Figure~\ref{fig:FPR_FNR} shows the FPR and FNR of the model as a function of the RB score threshold that is adopted to distinguish real from bogus sources.

\clearpage
\clearpage

\section{Pointing Details}\label{app:obs}
\startlongtable
\begin{deluxetable}{cccccc}
\centering
\tablecaption{{\bf Full observation log.} Columns include the band, RA and Dec at the centre of the pointing, 5$\sigma$ limiting magnitude, time (UTC) at the beginning of the exposure, and the source of reference images used in subsequent image differencing. Horizontal lines are used to distinguish different epochs for visibility.}\label{tab:details}
\tablehead{  
\colhead{band} & \colhead{RA (J2000)} & \colhead{Dec (J2000)} & \colhead{5$\sigma$ limiting} & \colhead{UTC} & \colhead{reference} \vspace{-2.5pt} \\
\colhead{} & \colhead{} & \colhead{} & \colhead{magnitude} & \colhead{} & \colhead{images} 
 }
\startdata
 \vspace{2pt}
$g$ & 00 42 00.00 & -23 30 00.00 & 23.0 & 2019-08-16T11:57:12.133 & PS1 3$\pi$ \\
$g$ & 00 54 00.00 & -23 30 00.00 & 22.9 & 2019-08-16T12:11:22.949 & PS1 3$\pi$ \\
$g$ & 00 50 00.00 & -27 30 00.00 & 22.7 & 2019-08-16T12:25:14.704 & PS1 3$\pi$ \\
$g$ & 00 46 00.00 & -21 30 00.00 & 22.4 & 2019-08-16T12:38:57.258 & PS1 3$\pi$ \\
$g$ & 00 58 00.00 & -27 30 00.00 & 22.4 & 2019-08-16T12:52:35.743 & PS1 3$\pi$ \\
$g$ & 00 51 27.69 & -23 56 38.4 & 22.6 & 2019-08-16T13:28:33.557 & PS1 3$\pi$ \\
$g$ & 00 46 41.12 & -23 56 38.4 & 22.9 & 2019-08-16T13:42:26.963 & PS1 3$\pi$ \\
$g$ & 00 53 00.31 & -25 04 58.2 & 22.9 & 2019-08-16T13:56:02.812 & PS1 3$\pi$ \\
$g$ & 00 48 07.44 & -25 04 41.1 & 22.9 & 2019-08-16T14:24:07.836 & PS1 3$\pi$ \\
$g$ & 00 54 37.59 & -26 17 02.5 & 23.0 & 2019-08-16T14:37:50.788 & PS1 3$\pi$ \\
$g$ & 00 49 47.57 & -26 15 24.4 & 23.1 & 2019-08-16T14:51:26.611 & PS1 3$\pi$ \\\hline
$g$ & 00 40 00.00 & -22 00 00.00 & 23.6 & 2019-08-21T11:03:45.415 & PS1 3$\pi$ \\
$g$ & 00 46 00.00 & -21 30 00.00 & 23.5 & 2019-08-21T11:17:21.039 & PS1 3$\pi$ \\
$g$ & 00 46 00.00 & -27 30 00.00 & 23.7 & 2019-08-21T11:31:11.689 & PS1 3$\pi$ \\
$g$ & 00 50 00.00 & -27 30 00.00 & 23.6 & 2019-08-21T11:45:11.843 & PS1 3$\pi$ \\
$g$ & 00 54 00.00 & -23 30 00.00 & 23.6 & 2019-08-21T11:58:47.234 & PS1 3$\pi$ \\
$g$ & 00 58 00.00 & -27 30 00.00 & 23.6 & 2019-08-21T12:12:31.620 & PS1 3$\pi$ \\
$g$ & 00 48 07.44 & -25 04 41.1 & 23.7 & 2019-08-21T13:45:30.243 & PS1 3$\pi$ \\
$g$ & 00 51 27.69 & -23 56 38.4 & 23.7 & 2019-08-21T13:59:06.066 & PS1 3$\pi$ \\
$g$ & 00 46 41.12 & -23 56 38.4 & 23.7 & 2019-08-21T14:12:41.501 & PS1 3$\pi$ \\\hline
$i$ & 00 42 00.00 & -23 30 00.00 & 22.1 & 2019-08-17T11:18:20.344 & PS1 3$\pi$ \\
$i$ & 00 54 00.00 & -23 30 00.00 & 21.0 & 2019-08-17T11:54:32.786 & PS1 3$\pi$ \\\hline
$i$ & 00 42 00.00 & -23 30 00.00 & 22.2 & 2019-08-18T10:35:25.063 & PS1 3$\pi$ \\
$i$ & 00 54 00.00 & -23 30 00.00 & 22.3 & 2019-08-18T10:49:00.903 & PS1 3$\pi$ \\
$i$ & 00 50 00.00 & -27 30 00.00 & 22.1 & 2019-08-18T11:02:52.961 & PS1 3$\pi$ \\
$i$ & 00 46 00.00 & -21 30 00.00 & 21.9 & 2019-08-18T11:16:51.396 & PS1 3$\pi$ \\
$i$ & 00 58 00.00 & -27 30 00.00 & 22.0 & 2019-08-18T11:30:55.731 & PS1 3$\pi$ \\
$i$ & 00 40 00.00 & -22 00 00.00 & 22.5 & 2019-08-18T11:44:41.983 & PS1 3$\pi$ \\
$i$ & 00 46 00.00 & -27 30 00.00 & 22.6 & 2019-08-18T11:58:31.068 & PS1 3$\pi$ \\
$i$ & 00 51 27.69 & -23 56 38.4 & 23.0 & 2019-08-18T12:12:11.272 & MegaCam \\
$i$ & 00 46 41.12 & -23 56 38.4 & 23.1 & 2019-08-18T12:25:46.275 & MegaCam \\
$i$ & 00 53 00.31 & -25 04 58.2 & 23.1 & 2019-08-18T12:39:22.686 & MegaCam \\
$i$ & 00 48 07.44 & -25 04 41.1 & 23.1 & 2019-08-18T12:52:57.308 & MegaCam \\
$i$ & 00 54 37.59 & -26 17 02.5 & 23.0 & 2019-08-18T13:06:37.296 & MegaCam \\
$i$ & 00 49 47.57 & -26 15 24.4 & 22.9 & 2019-08-18T13:20:11.331 & MegaCam \\\hline
$i$ & 00 54 00.00 & -23 00 00.00 & 22.7 & 2019-08-19T11:49:19.361 & MegaCam \\
$i$ & 00 54 00.00 & -26 50 00.00 & 22.7 & 2019-08-19T12:03:00.101 & MegaCam \\
$i$ & 00 50 00.00 & -26 50 00.00 & 22.7 & 2019-08-19T12:16:41.756 & MegaCam \\
$i$ & 00 44 48.00 & -24 00 00.00 & 22.8 & 2019-08-19T12:30:14.607 & MegaCam \\
$i$ & 00 51 27.69 & -23 56 38.4 & 22.7 & 2019-08-19T12:43:49.981 & MegaCam \\
$i$ & 00 46 41.12 & -23 56 38.4 & 22.7 & 2019-08-19T12:57:36.164 & MegaCam \\
$i$ & 00 53 00.31 & -25 04 58.2 & 22.8 & 2019-08-19T13:11:30.235 & MegaCam \\
$i$ & 00 48 07.44 & -25 04 41.1 & 22.8 & 2019-08-19T13:25:07.821 & MegaCam \\
$i$ & 00 54 37.59 & -26 17 02.5 & 22.7 & 2019-08-19T13:38:44.042 & MegaCam \\
$i$ & 00 49 47.57 & -26 15 24.4 & 22.7 & 2019-08-19T13:52:17.567 & MegaCam \\
$i$ & 00 56 00.00 & -24 00 00.00 & 22.8 & 2019-08-19T14:05:55.205 & MegaCam \\
$i$ & 00 50 00.00 & -23 30 00.00 & 22.8 & 2019-08-19T14:19:34.052 & MegaCam \\
$i$ & 00 46 00.00 & -25 30 00.00 & 22.9 & 2019-08-19T14:33:17.012 & MegaCam \\\hline
$i$ & 00 48 07.44 & -25 04 41.1 & 23.6 & 2019-08-21T12:27:43.381 & MegaCam \\\hline
$i$ & 00 48 07.44 & -25 04 41.1 & 24.0 & 2019-08-22T12:14:02.564 & MegaCam \\\hline
$i$ & 00 54 00.00 & -23 00 00.00 & 23.7 & 2019-08-23T11:25:11.107 & MegaCam \\
$i$ & 00 54 00.00 & -26 50 00.00 & 23.8 & 2019-08-23T11:39:10.051 & MegaCam \\
$i$ & 00 50 00.00 & -26 50 00.00 & 23.9 & 2019-08-23T11:52:45.866 & MegaCam \\
$i$ & 00 44 48.00 & -24 00 00.00 & 23.9 & 2019-08-23T12:06:20.488 & MegaCam \\
$i$ & 00 56 00.00 & -24 00 00.00 & 23.8 & 2019-08-23T12:20:03.561 & MegaCam \\
$i$ & 00 48 07.44 & -25 04 41.1 & 23.8 & 2019-08-23T12:33:49.717 & MegaCam \\
$i$ & 00 51 27.69 & -23 56 38.4 & 23.8 & 2019-08-23T12:47:26.802 & MegaCam \\
$i$ & 00 46 41.12 & -23 56 38.4 & 23.9 & 2019-08-23T13:01:06.660 & MegaCam \\
$i$ & 00 53 00.31 & -25 04 58.2 & 23.9 & 2019-08-23T13:15:56.624 & MegaCam \\
$i$ & 00 54 37.59 & -26 17 02.5 & 23.9 & 2019-08-23T13:29:58.428 & MegaCam \\
$i$ & 00 49 47.57 & -26 15 24.4 & 23.9 & 2019-08-23T13:43:46.399 & MegaCam \\
$i$ & 00 50 00.00 & -23 30 00.00 & 23.9 & 2019-08-23T13:57:37.187 & MegaCam \\
$i$ & 00 46 00.00 & -25 30 00.00 & 23.9 & 2019-08-23T14:11:11.481 & MegaCam \\\hline
$i$ & 00 50 00.00 & -26 50 00.00 & 23.8 & 2019-09-03T14:17:45.275 & \textit{N/A} \\
$i$ & 00 44 48.00 & -24 00 00.00 & 23.8 & 2019-09-03T14:31:16.199 & \textit{N/A} \\
$i$ & 00 50 00.00 & -23 30 00.00 & 23.7 & 2019-09-03T14:44:56.982 & \textit{N/A} \\
$i$ & 00 46 00.00 & -25 30 00.00 & 24.3 & 2019-09-04T12:01:03.927 & \textit{N/A} \\
$i$ & 00 48 07.44 & -25 04 41.1 & 23.7 & 2019-09-04T12:14:34.454 & \textit{N/A} \\
$i$ & 00 51 27.69 & -23 56 38.4 & 23.6 & 2019-09-04T12:28:04.990 & \textit{N/A} \\
$i$ & 00 46 41.12 & -23 56 38.4 & 23.8 & 2019-09-04T12:41:35.681 & \textit{N/A} \\
$i$ & 00 53 00.31 & -25 04 58.2 & 23.8 & 2019-09-04T12:55:07.055 & \textit{N/A} \\
$i$ & 00 54 37.59 & -26 17 02.5 & 23.8 & 2019-09-04T13:08:46.637 & \textit{N/A} \\
$i$ & 00 49 47.57 & -26 15 24.4 & 23.8 & 2019-09-04T13:22:16.274 & \textit{N/A} \\
$i$ & 00 54 00.00 & -23 00 00.00 & 23.8 & 2019-09-04T13:35:47.268 & \textit{N/A} \\
$i$ & 00 54 00.00 & -26 50 00.00 & 23.9 & 2019-09-04T13:49:20.905 & \textit{N/A} \\
$i$ & 00 56 00.00 & -24 00 00.00 & 23.8 & 2019-09-04T14:02:59.122 & \textit{N/A} \\\hline
$z$ & 00 48 07.44 & -25 04 41.1 & 22.7 & 2019-08-21T12:56:31.140 & DECaLS \\\hline
$z$ & 00 48 07.44 & -25 04 41.1 & 23.2 & 2019-08-22T12:42:50.901 & DECaLS \\
\enddata
\end{deluxetable}
\clearpage{}\clearpage{}

\onecolumngrid
\newpage
\section{Photometry for Selected TNS Sources}\label{app:TNS}
\movetabledown=3.55cm
\movetableright=0cm 
\tabletypesize{\scriptsize}
\begin{deluxetable*}{l|lllllllll}[!ht]
\centering
\tablecaption{\textbf{Photometry for TNS sources of interest.} $g_{1.7}$ denotes the $g$-band magnitude at 1.7 days, and so forth. All detections were made at $\geqslant5\sigma$ significance. The corresponding light curves for the sources are shown in Figure~\ref{fig:sample_lightcurves} (Section~\ref{sec:other-searches}). AT2019nbp was disqualified as a counterpart to GW190814 due to slow photometric evolution ($\Delta m < 0.1~\mathrm{mag/day}$, \citealt{andreoni20}) and a pre-merger detection (\citealt{ackley20}). AT2019noq and AT2019nxe have spectral classifications as Type II and Ia SNe, respectively (\citealt{andreoni20}). The host of AT2019ntm displayed a potential H$\alpha$ line which corresponded to $z=0.116$, outside the LVC $2\sigma$ confidence region of GW190814 (\citealt{ackley20}). }
\tablehead{
\colhead{TNS ID} & \colhead{$g_{1.7}$} & \colhead{$i_{3.7}$} & \colhead{$i_{4.7}$} & \colhead{$g_{6.6}$} & \colhead{$i_{6.6}$} & \colhead{$z_{6.6}$} & \colhead{$i_{7.7}$} & \colhead{$z_{7.7}$} & \colhead{$i_{8.7}$}  
}

\startdata
    AT2019nbp & $20.48 \pm 0.10$ & & $20.49 \pm 0.08$ & $20.78 \pm 0.10$ & & & & & \\                  
    AT2019noq & $21.89 \pm 0.10$ & $20.24 \pm 0.20$ & $20.28 \pm 0.18$ & $21.66 \pm 0.12$ & $19.93 \pm 0.24$ & $20.79 \pm 0.28$ & $19.92 \pm 0.21$ & $20.78 \pm 0.24$ & $20.19 \pm 0.11$ \\ 
    AT2019nxe & $22.33 \pm 0.02$ & $21.37 \pm 0.15$ & & $23.34 \pm 0.03$ & & & & & \\                  
    AT2019ntm & $21.98 \pm 0.06$ & $23.13 \pm 0.17$ & & $22.76 \pm 0.08$ & & & & & $23.92 \pm 0.10$\\ 
\enddata
\end{deluxetable*}\label{tab:tns}
{\ }

\end{document}